\documentclass{aa}

\usepackage{graphics,graphicx}
\usepackage{xcolor}
\usepackage{amsmath}
\usepackage{amssymb}
\usepackage{braket}
\usepackage{wrapfig}
\usepackage{caption}
\usepackage{subcaption}

\begin{document}
\title{Maser polarization through anisotropic pumping}
\titlerunning{Anisotropic pumping of masers}
\author{Boy Lankhaar\inst{1,2}, Gabriele Surcis\inst{3}, Wouter Vlemmings\inst{1}, Violette Impellizzeri\inst{2}}
\institute{\inst{1}Department of Space, Earth and Environment, Chalmers University of Technology, Onsala Space Observatory, 439 92 Onsala, Sweden; \\
\email{boy.lankhaar@chalmers.se} \\
\inst{2}Leiden Observatory, Leiden University, Post Office Box 9513, 2300 RA Leiden, Netherlands \\
\inst{3}INAF – Osservatorio Astronomico di Cagliari, Via della Scienza 5, 09047 Selargius, Italy 
}

\date{}

\abstract
{Polarized emission from masers is an excellent tool to study magnetic fields in maser sources. The linear polarization of the majority of masers is understood as an interplay of maser saturation and anisotropic pumping. However, for the latter mechanism, no quantitative modeling has been presented yet.}
{We aim to construct a comprehensive model of maser polarization, including quantitative modeling of both anisotropic pumping and the effects of maser saturation on the polarization of masers.}
{We extended regular (isotropic) maser excitation modeling with a dimension that describes the molecular population alignments, as well as including the linear polarization dimension to the radiative transfer. The results of the excitation analysis yielded the anisotropic pumping and decay parameters, which were subsequently used in one-dimensional proper maser polarization radiative transfer modeling.}
{We present the anisotropic pumping parameters for a variety of transitions from class I CH$_3$OH masers, H$_2$O masers, and SiO masers. SiO masers are highly anisotropically pumped due to them occurring in the vicinity of a late-type star, which irradiates the maser region with a strong directional radiation field. Class I CH$_3$OH masers and H$_2$O masers occur in association with shocks, and they are modestly anisotropically pumped due to the anisotropy of the excitation region.}
{Our modeling constitutes the first quantitative constraints on the anisotropic pumping of masers. We find that anisotropic pumping can explain the high polarization yields of SiO masers, as well as the modest polarization of unsaturated class I CH$_3$OH masers. The common $22$ GHz H$_2$O maser has a relatively weak anisotropic pumping; in contrast, we predict that the $183$ GHz H$_2$O maser is strongly anisotropically pumped. Finally, we outline a mechanism through which non-Zeeman circular polarization is produced, when the magnetic field changes direction along the propagation through an anisotropically pumped maser.}

\keywords{}

\maketitle
\section{Introduction}
Magnetic fields in maser sources can be studied by observing their polarized emission. While circular polarization yields information on the line-of-sight magnetic field strength, linear polarization yields information on the magnetic field direction. Polarized maser emission has been used to constrain magnetic field properties toward high-mass star-forming regions \citep[e.g.,~][]{vlemmings:08, vlemmings:10, lankhaar:18, surcis:22, surcis:23}, the circumstellar envelopes of evolved stars \citep{kemball:97, kemball:09, vlemmings:06, baudry:98}, as well as toward (ultra)luminous infrared galaxies \citep{robishaw:08}. Successful maser polarization observations have been performed for the molecules NH$_3$ \citep{mccarthy:23}, OH, H$_2$O, SiO, and CH$_3$OH. 

The phenomenon that underlies the polarization of masers is the Zeeman effect. The Zeeman effect is due to the (maser) molecule magnetic moment interacting with the magnetic field. Through the Zeeman effect, spectral lines are split into a manifold of transitions. In addition, the interaction of the molecular magnetic moment causes the molecule to precess around the magnetic field direction, thus endowing it with a preferred direction. The Zeeman splitting leads to the circular polarization of the maser line, which is on the order of a percent for non-paramagnetic maser species (e.g.,~H$_2$O, CH$_3$OH), but can be up to $100 \%$ for paramagnetic maser species (e.g.,~OH). On the other hand, due to the precession of the maser molecules around the magnetic field, the rate of stimulated emission (for a beamed maser) varies between the magnetic subtransitions, causing linear polarization of the maser emission, after the maser saturates \citep{goldreich:73}. Linear polarization can either be parallel or perpendicular to the magnetic field direction, depending on the angle between the magnetic field and the maser propagation direction. Numerical codes are available that are able to model the polarization of maser lines through the Zeeman effect and maser saturation \citep{lankhaar:19, tobin:23}.

There are a variety of mechanisms that can complement the production of maser polarization through the Zeeman effect and maser saturation \citep{western:83c, houde:14, wiebe:98, kylafis:83}. Most prominently, the phenomenon of anisotropic pumping is often invoked to explain high degrees of linear polarization in (mostly) SiO masers toward evolved stars \citep[e.g.,~][]{kemball:97, lankhaar:19}. When a maser is anisotropically pumped, the magnetic sublevels within a transition are pumped differently, thus resulting in a boost in the polarization. Anisotropically pumped masers can exhibit polarization when they are unsaturated  \citep{lankhaar:19}. The polarization direction is either parallel or perpendicular to the magnetic field direction, as long as the magnetic precession rate ($\sim$ s$^{-1}$/mG for a non-paramagnetic molecule) exceeds the rate of stimulated emission. However, since the initial hypothesis of anisotropic pumping by \citet{western:83c}, there has been no quantitative modeling of the anisotropic parameters, including their dependence on maser source properties, geometry and environment. %In the case that a maser is permeated by a magnetic field that changes direction along the amplification path, linear polarization may be converted to circular polarization \citep{wiebe:98}

For this work, we combined quantitative modeling of the anisotropic pumping of maser species with proper polarized maser radiative transfer modeling, to achieve a comprehensive model of maser polarization. We introduced a formalism to model the excitation of masers, that was set up as follows:
\begin{itemize}
    \item A multilevel maser excitation analysis was performed in a large velocity gradient (LVG) geometry. In contrast to regular maser excitation analyses, where it is common to assume all magnetic sublevels are populated equally, we relaxed this assumption and explicitly modeled the populations, also resolving the magnetic sublevels of quantum states. To keep the excitation problem tractable, we employed an irreducible tensor formalism \citep{landi:06, lankhaar:20a}, that afforded us to make approximations that reduced the computational time by orders of magnitude at the expense of minimal loss in accuracy \citep[$\lesssim\%$ ][]{lankhaar:20a}).
    \item The converged output of the excitation analysis includes the pumping and loss parameters for maser transitions. Since our treatment resolves the pumping and loss per magnetic sublevel in the maser states, these include the so-called anisotropic pumping and loss parameters. The stronger the pumping and loss terms vary within the magnetic sublevels of a maser quantum state, the stronger the associated maser transition is anisotropically pumped.
    \item The anisotropic pumping and loss parameters were extracted from the excitation analysis, and used in the radiative transfer code CHAMP \citep{lankhaar:19}, to perform a full polarized maser radiative transfer simulation. This yields observable parameters related to the maser polarization.
\end{itemize}
Using this model, we quantitatively model the anisotropic pumping and full maser polarization radiative transfer of different H$_2$O, class I CH$_3$OH and SiO masers.
%while also resolving the differently populated magnetic sublevels. To keep the excitation problem tractable, we employ an irreducible tensor formalism \citep{landi:06, lankhaar:20a}, that will afford us to make approximations that reduce the computational time by orders of magnitude at the expense of minimal loss in accuracy \citep[$\lesssim\%$ ][]{lankhaar:20a}). The maser excitation solutions are used as input to a polarized maser radiative transfer modeling simulation, using CHAMP \citep{lankhaar:19}.

This paper is structured as follows. In section 2, we present the formalism with which we performed our excitation modeling. In addition, we discuss ideal solutions to polarized radiative transfer of anisotropically pumped and Zeeman split masers, which will aid in the interpretation of the following simulations. In section 3, we describe the simulations that we performed on the polarization and anisotropic pumping of H$_2$O, CH$_3$OH, and SiO masers. In section 4, we discuss our simulations. We conclude in section 5.
\section{Theory}
\label{sec:theory}
In the following, we consider the excitation analysis of a molecule or atom that is embedded in a region with an anisotropic velocity gradient. Formally, due to the resulting anisotropic radiation field, one has to extend the excitation analysis by not only modeling the excitation of the molecular quantum state populations, but also their alignment elements. Aligned molecules produce polarized radiation, the radiative transfer of which has to be modeled accordingly. In section \ref{sec:exc}, we revise the relevant theory required to model the excitation analysis, including the alignment of the quantum states. This theory was first presented in \citet{goldreich:81} and \citet{deguchi:84}, but we restate it in an irreducible tensor formalism \citep{landi:06, lankhaar:20a}, that is computationally advantageous. In section \ref{sec:exc_maser}, focussing on maser transitions, we show how to extract the relevant anisotropic pumping parameters from an alignment resolved excitation analysis, and we present simple formulas to relate the polarization fraction of anisotropically pumped unsaturated masers to their anisotropic pumping parameters. In section \ref{sec:zee_pol} we compute the expected linear polarization fraction due to the Zeeman effect and in section \ref{sec:sat} we make some comments about the saturation and the magnetic saturation limits and how to relate these to the maser intensity and optical depth. 
\subsection{Excitation analysis}
\label{sec:exc}
\citet{goldreich:81} showed that when the velocity gradient is anisotropic, then this has as a consequence that the molecular population interacting with the emergent radiation field will be partially aligned. The alignment is strongest when collisional excitation is weak and the line optical depth is around unity, so that radiative interactions are strong, while also being optimally anisotropic. The alignment of the molecular population is either parallel or perpendicular to the magnetic field, provided that the magnetic precession rate, $g\Omega$, is always higher than rates of collisions or radiative events. We assume dominant magnetic precession for all non-masing transitions, which is an excellent assumption for conditions in interstellar gas \citep{goldreich:81, lankhaar:20a}. When a molecular population is partially aligned, the radiation that it emits will be accordingly polarized, either parallel or perpendicular to the magnetic field direction. \citet{goldreich:81} and later \citet{deguchi:84}, modeled the polarization of spectral lines excited in a plane parallel cloud. The geometry adopted by these authors was of a magnetic field either parallel or perpendicular to the velocity gradient. In the following, we expand on their modeling by formulating the alignment resolved excitation analysis in a computationally favorable formalism, while also considering arbitrary large-velocity gradient (LVG) geometries. Later, we use this formalism to model the excitation of astrophysical masers. 

To model the excitation properties of molecules embedded in an anisotropic radiation field, one needs to account for their alignment properties. We thus set out to set up the statistical equilibrium equations (SEE) for the molecular quantum state populations and their alignment. We follow \citet{landi:84} and \citet{lankhaar:20a} and formulate the molecular population in terms of their irreducible tensor elements, $\rho_{jk}$, where $j$ is the angular momentum and $k$ is the irreducible tensor rank \citep[see ][for a detailed discussion of population irreducible tensor elements]{landi:06}. The irreducible rank $k$ runs from $0$ to $2j$, and for our purposes, due to symmetry, can only assume even values. The element $k=0$ refers to the isotropic (or total) population of state $j$, while $k \geq 2$ elements refer to the alignment elements of quantum state $j$. The time-dependence of the irreducible tensor elements we note,
\begin{subequations}
\label{eq:SEE}
\begin{align}
\dot{\rho}_{jk} = \sum_{j' \neq j,\ k'} f_{jk;j'k'} \rho_{j'k'} - \sum_{k'} f_{jk;jk'} \rho_{jk'},
\end{align}
where the rate coefficients that describe radiative and collisional population events to quantum level $\ket{jk}$ from all other levels, for $j > j' $, are,
\begin{align}
f_{jk;j'k'} &= [j']B_{j'j}\sum_K r_{jk;j'k';K} \bar{J}_{K}^{jj'} + \sqrt{\frac{g_j}{g_{j'}}} C_{j'j} \delta_{kk'} \delta_{k0},
\end{align}
and for $j' > j $, they are,
\begin{align}
f_{jk;j'k'} &= [j']A_{j'j} r_{jk;j'k';K}^{\prime} + [j']B_{j'j}\sum_K r_{jk;j'k';K} \bar{J}_{K}^{jj'} \nonumber \\ &+ \sqrt{\frac{g_j}{g_{j'}}} C_{jj'} \delta_{kk'} \delta_{k0}.
\end{align}
The radiative and collisional depopulation rates from quantum level $\ket{jk}$ to all other levels are,
\begin{align}
f_{jk;jk'} &= \sum_{j' < j} A_{jj'} \delta_{k,k'} \nonumber \\ &+ \sum_{j \neq j'} \left[ [j]B_{jj'} \sum_K r_{jk;j'k';K}^{\prime \prime} \bar{J}_{K}^{jj'} + C_{jj'} \delta_{kk'} \right].
\end{align}
In the expressions for the rate-coefficients, we used the symbols $A_{j'j}$ for the Einstein coefficient for spontaneous emission and $B_{j'j}$ for the Einstein coefficients for absorption and stimulated emission, from level $j'$ to level $j$. The symbols $C_{j'j}$ denote the rate of collisional (de-)excitation from level $j'$ to level $j$. The square brackets indicate $[x]=2x+1$. The aligned populations are dependent on the isotropic and anisotropic radiation field elements, $\bar{J}_0^{jj'}$ and $\bar{J}_2^{jj'}$, which we define later on in Eqs.~(\ref{eq:rad_anis}). Furthermore, we introduced the coupling factors
\begin{align}
r_{jk;j'k';K} &= \sqrt{3[k,k',K]} \begin{Bmatrix}1 & j & j' \\ 1 & j & j' \\ K & k & k' \end{Bmatrix} \begin{pmatrix}k & k' & K \\ 0 & 0 & 0 \end{pmatrix}, \\
r_{jk;j'k';K}^{\prime} &= (-1)^{1+j+j'} \begin{Bmatrix}j & j & k \\ j' & j' & 1 \end{Bmatrix}, \\
r_{jk;j'k';K}^{\prime \prime} &= (-1)^{1+j-j'} \sqrt{3[k][k'][K]} \begin{Bmatrix}1 & 1 & K \\ j & j & j' \end{Bmatrix} \nonumber \\ &\times
\begin{Bmatrix}k & k' & K \\ j & j & j \end{Bmatrix} \begin{pmatrix}k & k' & K \\ 0 & 0 & 0 \end{pmatrix},
\end{align}
\end{subequations}
that encapsulate the angular momentum algebraic factors associated with radiative population events through absorption and stimulated emission and spontaneous emission events, and depopulation through absorption and stimulated emission events. The coupling factors are functions of Wigner-$3j$ -$6j$ and -$9j$ symbols, where the $6j$ and $9j$ symbols are denoted by curled brackets and have $6$ and $9$ elements, respectively, and the Wigner-$3j$ elements are denoted by round brackets.

\begin{figure*}[ht!]
    \centering
    \includegraphics[width=\textwidth]{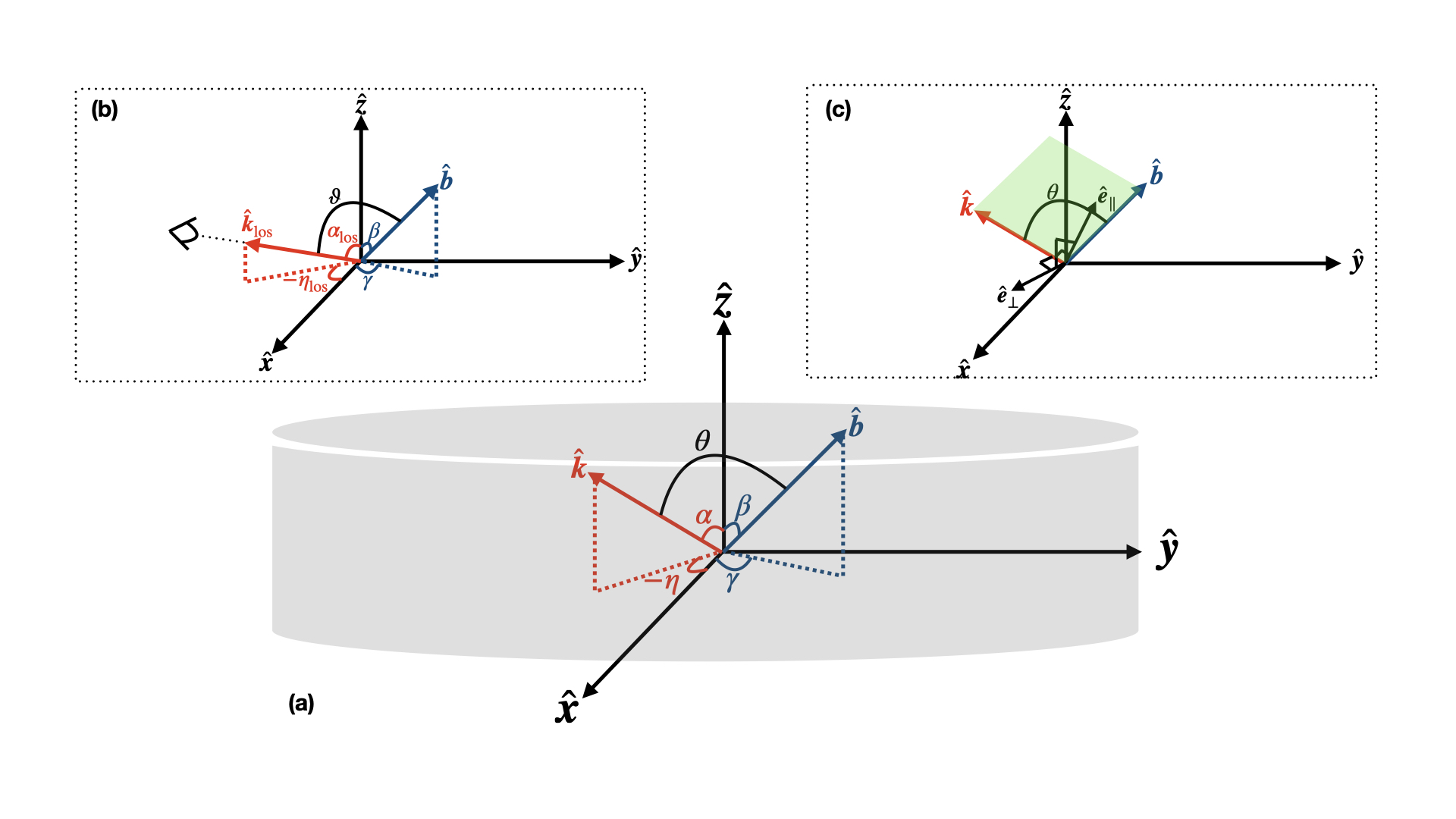}
    \caption{Coordinate system in the LVG geometry reference frame. LVG geometry is defined by the unit vectors, $\hat{{x}}$, $\hat{{y}}$ and $\hat{{z}}$. In (a), the magnetic field direction, $\hat{{b}}$, and radiation direction, $\hat{{k}}$, unit vectors are drawn, with their respective angles with respect to the LVG coordinate system. Throughout the paper, we use the symbol $\mu = \cos \theta = \hat{{b}} \cdot \hat{{k}}$, for the projection of the magnetic field onto the radiation direction. The translucent gray cylinder indicates the typical geometry we have used in our calculations, with weaker velocity gradients (and thus a larger LVG region) in the perpendicular direction. In (b), the special case when the radiation direction coincides with the line-of-sight is drawn. The projection angle of the magnetic field onto the line-of-sight direction is indicated by $\vartheta$. In (c), the magnetic field direction, $\hat{{b}}$, and radiation direction, $\hat{{k}}$, unit vectors are drawn, along with the unit vectors $\hat{e}_{\parallel}$ and $\hat{e}_{\perp}$, that define the radiation polarization directions, in, and perpendicular to, the plane (indicated in green) spanned by the magnetic field and radiation directions. In our calculations, we use an axisymmetric LVG geometry, with velocity gradient $\lambda$ in the $\hat{{z}}$-direction, that defines the symmetry axis, and velocity gradient $\lambda / \epsilon$ in the $\hat{{x}}$ and $\hat{{y}}$ directions. Unless otherwise indicated, we used $\beta = 0$ and $\vartheta = \pi/2$ for the magnetic field and line-of-sight direction in our calculations.} 
    \label{fig:geometry}
\end{figure*}

We proceed to consider the radiative transfer, where our ultimate goal is to acquire expressions for the radiation field elements, $\bar{J}_0^{jj'}$ and $\bar{J}_2^{jj'}$. In Fig.~(1), we give a sketch of the vectors that are relevant to the radiative transfer in our problem. We consider the radiative transfer in direction, $\hat{k}$, that is described by the angles $\Omega$ ($[\alpha\eta]$, when defined in the LVG reference frame, see Fig.~1). The partial alignment of the molecular medium entails a different rate of interaction with the polarization components of the radiation field. Because the molecular alignment is either parallel or perpendicular to the magnetic field, it is most convenient to express the radiation field with polarization components in the directions, $\hat{e}_{\parallel}$ and $\hat{e}_{\perp}$, parallel and perpendicular to the rejection of the magnetic field direction from the radiation direction (see Fig.~\ref{fig:geometry}c). The polarization components of the radiation field specific intensity in the parallel and perpendicular directions are indicated by $I_{\parallel}$ and $I_{\perp}$, respectively. The radiative transfer equation of both polarization modes of the radiation field near the resonant frequency of the transition $j\to j'$, can then be expressed as \citep{goldreich:81, lankhaar:20b},
\begin{subequations}
\label{eq:rad_trans}
\begin{align}
\frac{d}{ds_{\Omega}} I_{\parallel,\perp} (\nu,\Omega) = -\kappa_{\parallel,\perp} (\nu,\Omega) I_{\parallel,\perp}(\nu,\Omega) + \epsilon_{\parallel,\perp} (\nu,\Omega),
\end{align}
where,
\begin{align}
\kappa_{\parallel,\perp} (\nu,\Omega) &=        [k_0 + k_2 P_{\parallel,\perp}(\Omega)]\phi_{\nu-\nu_{jj'}} \\
\epsilon_{\parallel,\perp} (\nu,\Omega) &= [e_0 + e_2 P_{\parallel,\perp}(\Omega) ]\phi_{\nu-\nu_{jj'}},
\end{align}
where $\phi_{\nu-\nu_{jj'}}$ is the line profile, centered around the resonance frequency, $\nu_{jj'}$, and where the alignment propagation constants are defined,
\begin{align}
k_0 &= \frac{h \nu}{4\pi} [j]B_{jj'} \sqrt{3} (-1)^{j+j'} \nonumber \\ &\times \left[ \begin{Bmatrix}1 & 1 & 0 \\ j & j & j'\end{Bmatrix} \rho_{j0} -  \begin{Bmatrix}1 & 1 & 0 \\ j'& j'& j \end{Bmatrix} \rho_{j'0} \right],\\
k_2 &= \frac{h \nu}{4\pi} [j]B_{jj'}\sqrt{3} (-1)^{j+j'} \nonumber \\ &\times\left[ \begin{Bmatrix}1 & 1 & 2 \\ j & j & j'\end{Bmatrix} \rho_{j2} -  \begin{Bmatrix}1 & 1 & 2 \\ j'& j'& j \end{Bmatrix} \rho_{j'2} \right], \\
e_0 &= \frac{h \nu}{4\pi} A_{j'j}\sqrt{3} (-1)^{j+j'} \begin{Bmatrix}1 & 1 & 0 \\ j'& j'& j \end{Bmatrix} \rho_{j'0}, \\
e_2 &= \frac{h \nu}{4\pi} A_{j'j}\sqrt{3} (-1)^{j+j'} \begin{Bmatrix}1 & 1 & 2 \\ j'& j'& j \end{Bmatrix} \rho_{j'2},
\end{align}
and their angular weights are,
\begin{align}
P_{\parallel} (\Omega) &= \frac{3\mu^2 - 2}{\sqrt{2}}, \\
P_{\perp} (\Omega) &= \frac{1}{\sqrt{2}},
\end{align}
\end{subequations}
where we defined $\mu = \hat{b} \cdot \hat{k}$ (see Fig.~\ref{fig:geometry}).
We note that Eqs.~(\ref{eq:rad_trans}) reduce to the usual line radiative transfer equation \citep[see, e.g.,~][]{rybicki:08} when no alignment is present in either of the transition energy levels: $k_2=e_2=0$. 

We now return to the SEE. We noted the dependence of the radiative interactions on the radiation (alignment) elements, which we may express in terms of the polarization elements \citep{goldreich:81, landi:06}
\begin{subequations}
\label{eq:rad_anis}
\begin{align}
\bar{J}_{0}^{jj'} &=  \int \frac{d \Omega}{4\pi} \int d\nu \ \phi_{\nu - \nu_{jj'}} [I_{\parallel} (\nu,\Omega) + I_{\perp}(\nu,\Omega)] \\
\bar{J}_{2}^{jj'} &=  \int \frac{d \Omega}{4\pi} \int d\nu \ \phi_{\nu - \nu_{jj'}} [I_{\parallel}(\nu,\Omega) P_{\parallel}(\Omega) + I_{\perp}(\nu,\Omega) P_{\perp}(\Omega)].
\end{align}
Using the LVG approximation, with the velocity gradient $\lambda (\Omega)$ in direction $\Omega$, it can be shown that
\begin{align}
\int d\nu \ \phi_{\nu - \nu_{jj'}}  I_{\parallel,\perp} (\nu,\Omega) &= \frac{\epsilon_{\parallel,\perp}}{\kappa_{\parallel,\perp}}\left[1 - \beta_{\parallel,\perp}(\Omega)\right] \nonumber \\  &+  I_{\parallel,\perp}^{(0)}(\nu,\Omega)\beta_{\parallel,\perp}(\Omega) , \\
\beta_{\parallel,\perp}(\Omega) &= \frac{1 - e^{-\tau_{\parallel,\perp}(\Omega)}}{\tau_{\parallel,\perp}(\Omega)}, \\
\tau_{\parallel,\perp}(\Omega) &= \frac{c k_{\parallel,\perp} (\Omega)}{\nu \lambda (\Omega)},
\end{align}
\end{subequations}
where we have denoted the background radiation field of the particular transition by $I_{\parallel,\perp}^{(0)}(\nu,\Omega)$. The expressions of Eqs.~(\ref{eq:rad_anis}) relate the radiation (alignment) elements that appear in the SEE of Eqs.~(\ref{eq:SEE}) to the population (alignment) elements, $\rho_{jk}$. The LVG approximation has thus afforded us to formulate the SEE as a set of (nonlinear) equations that may be solved with root-finding algorithms such as the Newton-Raphson method. The solution yields the (alignment) populations of a set of quantum states $\ket{jk}$. The (alignment) populations are coupled through radiative transitions, characterized by the Einstein coefficients and their associated frequencies, collisional transitions, whose rates are dependent on the number density of the collision partner. The radiative transitions are dependent on the LVG geometry, characterized by the (direction dependent) velocity gradient, $\lambda(\Omega)$, and the background radiation field $I_{\parallel,\perp}^{(0)}(\nu,\Omega)$. 

\subsection{Anisotropic maser pumping}
\label{sec:exc_maser}
For some molecules, particular transitions, under specific conditions, become population inverted, and their radiative transfer is characterized by exponential amplification due to the dominant process of stimulated emission. These transitions are referred to as masers (Microwave Amplified by Stimulated Emission of Radiation) and occur for a variety of molecules toward different astrophysical objects. The excitation of maser transitions are often effectively modeled using an analysis that employs the LVG approximation \citep{elitzur:92, gray:12}, but commonly such analyses lack any consideration of the alignment of the involved quantum states. In the next section, we consider excitation of maser transitions using the LVG approximation, while also modeling the associated alignment of the maser states, making use of the formalism described above. In anticipation of this, in the remainder of this subsection, we outline how to, (i) extract the anisotropic pumping parameters of a maser from the results of an (alignment resolved) LVG calculation, and (ii) apply these results to the polarized radiative transfer of anisotropically pumped unsaturated masers.

Maser transitions exhibit exceptionally strong radiation fields, which are often beamed. The exponential amplification of radiation tends to invalidate the local approximation that is part of the LVG approximation. Therefore, it is useful to consider the maser radiative transfer and the excitation of the maser levels in isolation from the rest of the molecular transitions and levels. We consider a maser transition, where we let the upper and lower state of the maser transition be $j_a$ and $j_b$. We separate in the excitation analysis the states that are involved in the maser transitions, and all other states, which we refer to as the reservoir. We separate the SEE in these terms, and because of the dominance of the stimulated emission in a maser transition, we ignore spontaneous emission events and collisional transitions between maser states in the excitation analysis. We then note the time-dependence of the maser states
\begin{subequations}
\label{eq:maser_excitation}
\begin{align}
\dot{\rho}_{j_a k} &= 0 = \lambda_{j_a k} - \sum_{k'} \left[\gamma_{j_a k k'} \rho_{j_a k'} + [j_a] B_{j_a j_b}\right. \nonumber \\ &\times \left.  \sum_K \left(r_{j_a k;j_b k';K} \rho_{j_b k'} - r_{j_a k,j_b k';K}^{\prime \prime} \rho_{j_a k'} \right) \bar{J}_{K}^{j_a,j_b} \right], \\
\dot{\rho}_{j_b k} &= 0 = \lambda_{j_b k} - \sum_{k'} \left[\gamma_{j_b k k'} \rho_{j_b k'} -  \right. [j_a] B_{j_a j_b} \nonumber \\ &\times \left. \sum_K \left(r_{j_b k;j_a k';K}^{\prime \prime} \rho_{j_b k'} - r_{j_b k,j_a k';K} \rho_{j_a k'} \right) \bar{J}_{K}^{j_a,j_b} \right],
\end{align}
where we have defined the pumping and decay operators
\begin{align}
\lambda_{j_a k} &= \sum_{j' \neq j_a,j_b} f_{j_a k; j' k'} \rho_{j' k'}, \\
\lambda_{j_b k} &= \sum_{j' \neq j_a,j_b} f_{j_b k; j' k'} \rho_{j' k'}, \\
\gamma_{j_a k k'} &= \sum_{\substack{j' < j_a \\ j' \neq j_b}} A_{j_a j'} \delta_{kk'} \nonumber \\ 
&+ \sum_{j' \neq j_a,j_b} \left[ [j_a] B_{j_a j'} \sum_K r_{jk;j'k';K}^{\prime \prime} J_K^{j_a j'}  +  C_{j_a j'} \delta_{kk'} \right].
\end{align}
The pumping operators, $\lambda_{jk}$, describe the populating interactions from the reservoir to the maser levels, while the decay operators, $\gamma_{j k k'}$, describe the depopulating interactions from the maser levels to the reservoir. In an anisotropic excitation region, it is possible for both the pumping and the decay operators to have significant alignment terms. Then, it is easily recognized, that for such a maser, the $\rho_{j_a/j_b 2}$ alignment populations are significant and polarized maser emission is produced. The pumping and decay operators are easily computed from the converged output of an excitation analysis described in the previous subsection. 

We assume that the maser radiation field is significantly beamed, thus being approximately one-dimensional and subtending a small solid angle, $\Delta \Omega$, in direction $\Omega_m$, so that
\begin{align}
\bar{J}_{0}^{j_aj_b} &=  \frac{\Delta \Omega}{4\pi} \int d\nu \ \phi_{\nu - \nu_{j_aj_b}} [I_{\parallel} (\nu,\Omega_m) + I_{\perp}(\nu,\Omega_m)] \\
\bar{J}_{2}^{j_aj_b} &=  \frac{\Delta \Omega}{4\pi} \int d\nu \ \phi_{\nu - \nu_{j_a j_b}} \nonumber \\ &\times [I_{\parallel}(\nu,\Omega_m) P_{\parallel}(\Omega_m) + I_{\perp}(\nu,\Omega_m) P_{\perp}(\Omega_m)].
\end{align}
\end{subequations}
Assuming that the maser is beamed, and extracting the pumping and decay operators from an excitation analysis, a proper polarized maser radiative transfer propagation can be performed using Eqs.~(\ref{eq:maser_excitation}) in conjunction with Eq.~(\ref{eq:rad_trans}).

%To first order in $J_2^{j,j'}/J_0^{j,j'}$, for all transitions, and for a weak maser field, we have the populations relevant to the propagation
%\begin{subequations}
%\begin{align}
%\rho_{j_a 0} &= \frac{\lambda_{j_a 0}}{\gamma_{j_a00}}, \qquad
%\rho_{j_a 2} = \frac{\lambda_{j_a 2} - \lambda_{j_a 0} %\frac{\gamma_{j_a20}}{\gamma_{j_a00}}}{\gamma_{j_a22}}, \\
%\rho_{j_b 0} &= \frac{\lambda_{j_b 0}}{\gamma_{j_b00}}, \qquad 
%\rho_{j_a 2} = \frac{\lambda_{j_b 2} - \lambda_{j_b 0} %\frac{\gamma_{j_b20}}{\gamma_{j_b00}}}{\gamma_{j_b22}}. 
%\end{align}
%\end{subequations}
For an unsaturated maser, the $\bar{J}_K^{j_aj_b}$-dependent terms are negligible in determining the (alignment) populations $\rho_{j_a k}$ and $\rho_{j_b k}$. The populations are then easily extracted from only the pumping and decay terms, using Eqs.~(\ref{eq:maser_excitation}). As long as the maser remains unsaturated, the (alignment) populations remain approximately constant and the radiative transfer equations of Eqs.~(\ref{eq:rad_trans}) may be solved analytically. The linear polarization fraction we define as 
\begin{align}
\label{eq:linpolfrac}
p_Q = \frac{I_{\parallel}-I_{\perp}}{I_{\parallel}+I_{\perp}}.
\end{align}
Neglecting the effects of spontaneous emission, the polarization fraction of an anisotropically pumped unsatured maser is,
\begin{align}
\label{eq:pQ_anis}
p_Q^{\mathrm{anis}} &\simeq \tanh{ \left[k_2^{(0)} s \phi_{\nu} \frac{3 \sin^2 \vartheta}{2\sqrt{2}} \right]} \nonumber \\ &= \tanh{ \left[ -\tau_\nu^{(0)} q_{\mathrm{anis}} \sin^2 \vartheta \right]},
%BL 6/01
\end{align}
where $s$ is the maser length, $\tau_{\nu}^{(0)}$ the maser optical depth (positive when population inverted) and $\cos \vartheta$ is the projection of the maser radiation field direction onto the magnetic field direction. We defined the anisotropic pumping factor $q_{\mathrm{anis}} = -\frac{3k_2^{(0)}}{2\sqrt{2}k_0^{(0)}}$, where the $(0)$-superscripts denote that these are the propagation coefficients in the unsaturated limit. It should be noted that $k_2^{(0)}$, and thus $q_{\mathrm{anis}}$, may assume a positive or negative value. In case $q_{\mathrm{anis}}$ is positive, the polarization is oriented perpendicular to the projected magnetic field direction, while in case it is negative, the polarization is oriented along the projected magnetic field direction. In the limit of weakly polarized masers, the polarization scales linearly with the maser optical depth:
\[
p_Q^{\mathrm{anis}} \simeq  -\tau_\nu^{(0)} q_{\mathrm{anis}} \sin^2 \vartheta .
\]
%The maximum optical depth of the maser we may extract from the maximum Sobolev length of the transition in the region: $\tau_{\mathrm{max}} = k_0 c/\nu_0 \lambda_{\mathrm{min}}$.

\subsection{Unsaturated maser limit}
\label{sec:sat}
Equation~(\ref{eq:pQ_anis}) is valid for unsaturated masers. Masers are considered to be unsaturated, when the rate of stimulated emission, induced by the maser radiation, is lower than the maser decay rate \citep{elitzur:91},
\[
B_{j_a,j_b} I_{\nu_0} \frac{\Delta \Omega}{4\pi} \ll \bar{\gamma},
\]
where $\bar{\gamma}=(\gamma_{j_a 00}+\gamma_{j_b00})/2$. Considering the relation between the maser optical depth and the linear polarization fraction in unsaturated masers, it will be helpful to compute the optical depth at which the maser saturates: the saturation optical depth. We assume that in the unsaturated maser regime, the maser radiation intensity amplifies a background radiation field, $I_0$, exponentially $I_{\nu_0} \simeq I_0 e^{\tau_{\nu_0}}$. The background radiation field may either be the ambient background radiation field, or it may emerge from the maser population itself. The latter scenario is relevant for sources where the maser excitation temperature, $|T_{\mathrm{exc}}|$, exceeds the ambient background radiation temperature. We take the maser solid angle, $\Delta \Omega/4\pi = \frac{0.05\  \mathrm{sr}}{\tau_{\nu_0}}$, as inversely proportional to the optical depth\footnote{This relation follows from maser excited in a plane-parallel with a velocity-gradient aspect ratio of $\epsilon=0.01$.}, one may solve 
\begin{align}
I_{\nu_0} \frac{\Delta \Omega}{4\pi} &= \bar{\gamma}/B_{j_a,j_b} \nonumber \\
&= \frac{I_0 e^{\tau_{\mathrm{sat}}}\times 0.05\ \mathrm{sr} }{\tau_{\mathrm{sat}}},
\end{align}
to obtain the saturation optical depth $\tau_{\mathrm{sat}}$.

When the stimulated emission rate approaches and exceeds the maser decay rate, the interaction of the maser states with the maser radiation fields starts to become important. In Eqs.~(\ref{eq:maser_excitation}),  this may be recognized through that the $J_K^{j_a j_b}$-dependent terms have become significant past saturation limit. This has as a consequence that the maser amplification tends to become linear, but also that the maser state alignment and subsequently the maser polarization, are affected by the interaction of the maser states with the directional maser radiation field. The polarization behavior of such a maser is then a hybrid between the anisotropic pumping (and/or Zeeman polarization) and the regular polarization of a saturated maser \citep[see, e.g.,~][]{lankhaar:19}. This may either enhance or diminish the predicted polarization fraction from the anisotropic pumping alone. The polarization produced past the saturation limit remains oriented parallel or perpendicular to the projected magnetic field direction, provided that the rate of stimulated emission induced by the maser radiation, is lower than the magnetic precession rate
\[
B_{j_a,j_b} I_{\nu_0} \frac{\Delta \Omega}{4\pi} \ll g \Omega.
\]
According to the one-dimensional maser propagation equation, $I_{\nu_0} \Delta \Omega$ grows linearly with the (unsaturated) optical depth after the saturation optical depth has been reached \citep{elitzur:91}. We thus may compute the magnetic saturation optical depth as,
\begin{align}
\tau_{\mathrm{mag.~sat}} = \tau_{\mathrm{sat}}+\frac{g\Omega}{\bar{\gamma}}.
\end{align}
When the magnetic saturation limit is approached, the symmetry axis of the maser molecules is rotated from being along the magnetic field direction, to being along the maser radiation field direction. The rotation of the symmetry axis is associated with the production of linearly polarized radiation that is not oriented with respect to the magnetic field. The rotation of the linear polarization and the molecular symmetry axis is additionally associated with the production of high degrees of circular polarization. 

\section{Simulations}
The excitation analyses were performed by solving the set of nonlinear equations described in Eqs.~(\ref{eq:SEE}-\ref{eq:rad_anis}), under the physical constraint $\sum_j [j]^{1/2} \rho_{j0} = n_{\mathrm{mol}}$, where $n_{\mathrm{mol}}$ is the number density of the molecule of interest. Since the solutions to the radiative transfer equations of Eq.~(\ref{eq:rad_trans}) are divergent in the case of population inversion, due to the local approximation, we put the escape probability of these transitions at unity to later solve the (polarized) radiative transfer in isolation. This is an excellent approximation for H$_2$O and CH$_3$OH, where maser transitions are isolated from each other, and only connected to the reservoir \citep{neufeld:87, neufeld:94b}. For SiO, this approximation is not as good, as SiO maser transitions occur consecutively in the $J\to J-1$ transitions. For example, the radiative transfer of the $J \to J-1$ and $J+1 \to J$ transitions are directly linked, since they both involve level $J$. If both these transitions occur as a maser, and any of them is saturated, then they cannot strictly be treated in isolation \citep{lockett:92}. For non-maser transitions, solutions to the radiative transfer equations of Eq.~(\ref{eq:rad_trans}) are a function of the direction-dependent velocity gradient $\lambda (\Omega)$. For our calculations, we considered an axisymmetric system, exhibiting a velocity gradient $\lambda_{\parallel} = \lambda$ along the symmetry axis and a velocity gradient $\lambda_{\perp} = \lambda / \epsilon$ perpendicular to the symmetry axis (see Fig.~1). Thus, geometries with $\epsilon < 1$ have a correspondence to filamentary type geometries, and $\epsilon > 1$ to disk type geometries \citep{elitzur:89}. In our simulations, we assumed the magnetic field direction to be along the symmetry axis, $\hat{b} = \hat{z}$, which is computationally advantageous, but our formalism may be used for arbitrary magnetic field directions. This allows us to relate the projection, $\mu = \hat{z} \cdot \hat{k} = \hat{b} \cdot \hat{k} $, and note the general velocity gradient
\[
\lambda (\Omega) = \lambda (\mu) = \lambda \left( \mu^2 + \frac{1-\mu^2}{\epsilon} \right).
\]
We considered disk-like geometries in our calculations, adopting an aspect ratio of $\epsilon=10$. The LVG optical depths, $\tau_{\mathrm{LVG}} = \frac{c k}{\nu \lambda} \propto \frac{N_{\mathrm{mol}}}{\Delta v_{\mathrm{FWHM}}}$, are proportional to the specific column density \citep[see, e.g.,~][]{hollenbach:13}, which we take as a general input parameter of our simulations.\footnote{In the literature on collisionally pumped masers, the parameter $\xi \propto \frac{n_{\mathrm{H}_2} N_{\mathrm{mol}}}{\Delta v_{\mathrm{FWHM}}}$ is often used to characterize and analyze the results of excitation analyses \citep{elitzur:89}. It can be shown that in the limit of optically thick transitions, then the rate equations depend only on the $\xi$-parameter and the temperature. However, the optically thick limit required for this degeneracy to emerge is not perfectly fulfilled \citep[see, e.g., figure 6 of][and note the number density dependence of the results]{neufeld:91}.  Also, the alignment of quantum states critically depends on collisional rates and accordingly the gas number density. We therefore maintain to characterize our simulations in terms of gas number density and specific column density.}
We always report the specific column density along the axis with the highest velocity gradient, so that it is easily compared to regular plane parallel slab LVG calculations. Collisional rate coefficients depend on the temperature and number density of the main collision partner (H$_2$) which we also take as general input parameters. In all calculations, we neglected alignment states above $k=2$. The relative tolerance for convergence of the excitation solution was set to $10^{-6}$.

The results of the excitation analyses are reported using the quantities, $\tau_{\mathrm{m}}^{\perp}$, which is the maser optical depth along the short axis, $q_{\mathrm{anis}}$, which is a parameter describing the pumping anisotropy, and is defined in Eq.~(\ref{eq:pQ_anis}), the maser decay rate, $\bar{\gamma}$, and the excitation temperature, $T_{\mathrm{exc}}$. The excitation temperature is a proxy of the relative inversion of the maser levels, and is defined, for a transition between upper level $a$ and lower level $b$, at frequency $\nu$, and with level degeneracies $g_a$ and $g_b$,
\begin{align}
e^{h\nu / k T_{\mathrm{exc}}} = \sqrt{\frac{g_b}{g_a}} \frac{\rho_{j_{a0}}}{\rho_{j_{b0}}},
\end{align}
where it should be noted that we opted to report a positive excitation temperature, even though the maser is population inverted.
%, leading to an order $<\%$ error in the alignment populations. %The Einstein coefficients, radiative transition frequencies and collisional rate coefficients were inputted as constants. Later, we specify per molecule where we have taken these constants from. 
%We computed the alignment of molecular states assuming initially the alignment axis to be along the symmetry axis. The alignment with respect to the magnetic field direction, $\hat{b}$, may subsequently be computed through
%\[
%\left[ \frac{\rho_{j2}}{\rho_{j0}}\right]_{\hat{b}} = \frac{3 \left(\hat{b}\cdot \hat{r}\right)^2 - 1}{2}\left[ \frac{\rho_{j2}}{\rho_{j0}}\right]_{\hat{r}}.
%\]

\subsection{H$_2$O masers}
H$_2$O masers can be excited under various conditions. Most commonly, H$_2$O masers occur in association with shocked material, where post-shock densities are enhanced to $n \sim 10^9 \ \mathrm{cm}^{-3}$ and the gas is heated to $T \gtrsim 400 \ \mathrm{K}$  \citep{hollenbach:13}. Shock excited H$_2$O masers occur in (high-mass) star-forming regions \citep{gray:12}, and can appear in association with fast collimated outflows emerging from late-type stars \citep{imai:07}. H$_2$O masers can also arise in spherical shells, at around $10$ stellar radii, toward evolved stars, where their excitation is significantly affected by the ambient infrared radiation field and warm dust \citep{gray:16,gray:22}. Extragalactic H$_2$O megamasers are understood to emerge in the vicinity of active galactic nuclei (AGNs). There, they occur either in the inner regions ($<1$ pc) of the accretion disk \citep{miyoshi:95, gao:16}, or in association with the jet that is launched from the AGN \citep{peck:03, sawada:08, gallimore:23}. Disk masers occur in association with an X-ray dissociation region \citep{neufeld:94b, collison:95}, while jet masers are thought to be excited in shocked gas. In either case, densities and temperatures are similar to those of the shocked regions in which galactic H$_2$O masers occur, but their size extent is much larger. 

 %To model megamasers, one therefore puts the range of column densities two orders of magnitude greater. For such large column densities, it is important to also account for the contribution of the dust to the radiative transfer, as otherwise the radiative transitions would saturate and population inversion is suppressed \citep{collison:95}. Unfortunately, due to the reliance of the LVG approximation on a line profile, combining it with a continuum absorption contribution is associated with a steep increase in computational costs. \citet{hollenbach:79} outlined an approach to include continuum absorption in the photon escape probability, but their approximations are only valid as isotropic escape probabilities and their anisotropic information is lost. Due to the computational challenges, we will not model megamasers specifically, but rather use the estimates of shock-excited H$_2$O masers and discuss a scale-up to megamaser excitation.

We investigated H$_2$O masers in shocked geometries. In the discussion section, we dedicate some attention to extrapolating our results to the other types of H$_2$O maser excitation. We studied the $22$ GHz transition, which is the strongest and most widely studied H$_2$O maser transition, and also the (sub)millimeter maser transitions around $183$ GHz and $321$ GHz. We studied ortho- and para-H$_2$O masers seperately. For each symmetry species, we modeled the excitation and (anisotropic) pumping parameters using the $45$ lowest rotational levels in the ground vibrational state. Radiative and collisional rates were taken from the LAMDA database \citep{schoier:05}, where we used the collisional rates of \citet{daniel:11}. We modeled the maser region in a disk-like LVG geometry, adopting a modest aspect ratio of $\epsilon=10$. As conditions generally representative of H$_2$O masers, we consider the excitation of H$_2$O masers at number density $n_{\mathrm{H}_2} = 10^{9}\ \mathrm{cm}^{-3}$ and temperature of $T=1000$ K, as well as H$_2$O masers at $n_{\mathrm{H}_2} = 10^{8}\ \mathrm{cm}^{-3}$ and temperature of $T=400$ K for a range of specific column densities. We assumed an isotropic background radiation field of $T_{\mathrm{B}} = 2.73$ K.
\begin{figure*}[ht!]
\centering
    \begin{subfigure}[b]{0.4\textwidth}
      \includegraphics[width=\textwidth]{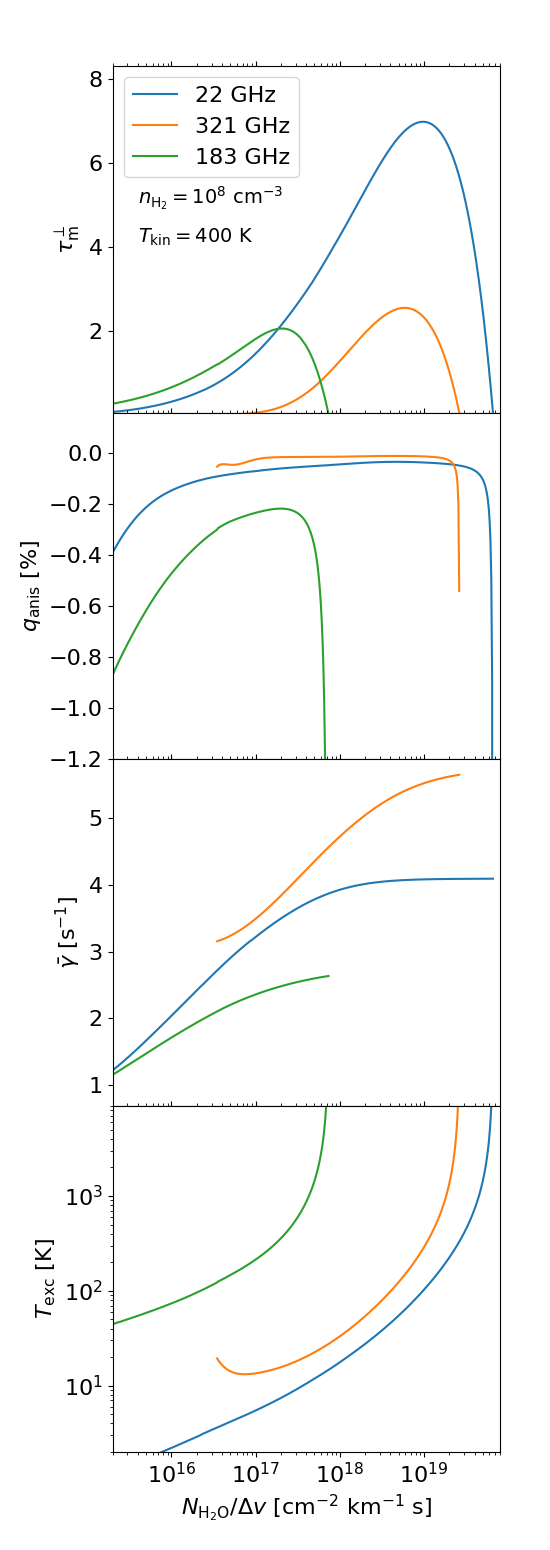}
      \caption{}
      \label{fig:water_400}
    \end{subfigure}
%\hfill
    \begin{subfigure}[b]{0.4\textwidth}
      \includegraphics[width=\textwidth]{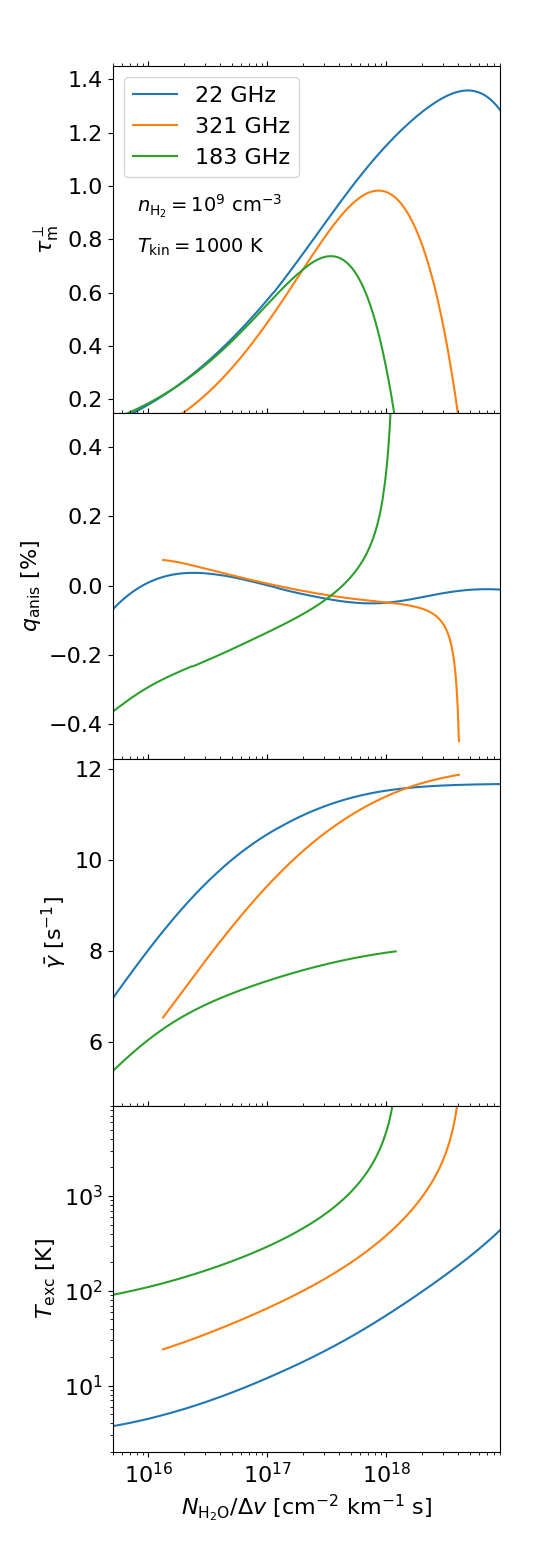}
      \caption{}
      \label{fig:water_1000}
    \end{subfigure}
\caption{Plots of the maser (anisotropic) pumping parameters of H$_2$O maser transitions as a function of the specific column density. Excitation conditions are indicated inside the figure. Plots are given for the maser optical depth (upper row), anisotropic pumping factor (second row), decay rate (third row) and excitation temperature (lower row) of common maser transitions at $22$ GHz, $321$ GHz and $183$ GHz.} 
\label{fig:water_standard}
\end{figure*}

Results of the maser excitation modeling of H$_2$O masers are given in Fig.~(\ref{fig:water_standard}). Maser action is predicted for all investigated transitions. For all transitions, the maser optical depth, $\tau_{\mathrm{m}}^{\perp}$, that is measured by the optical depth along the short axis of the maser-disk, is a strong function of the specific column density, slowly rising until it reaches maximal inversion, where after a precipitous fall in population inversion occurs and the levels become thermalized. Strongest maser action is expected at lower densities, but we should note that maximal population inversion occurs, for the $22$ GHz and $321$ GHz transitions, at specific column densities on the order of $N_{\mathrm{H_2O}}/\Delta v \sim 10^{19} \mathrm{cm}^{-2}\ \mathrm{km}^{-1} \ \mathrm{s}$, that for $n_{\mathrm{H}_2}=10^8 \ \mathrm{cm}^{-3}$ gas corresponds to maser sizes of $d \sim 10^{15} \ \mathrm{cm}$, assuming $x_{\mathrm{H_2O}} \sim 10^{-4}$. Such large masers are not expected \citep{hollenbach:13}. Maximal $183$ GHz emission in $n=10^8\ \mathrm{cm}^{-3}$ gas, is expected at physically reasonable maser sizes, where the $22$ GHz is expected to be significantly pumped as well. For higher density and temperature gas, significant maser optical depths are expected for all investigated masers at $N_{\mathrm{H_2O}}/\Delta v \gtrsim 10^{17} \mathrm{cm}^{-2}\ \mathrm{km}^{-1} \ \mathrm{s}$, with the $22$ GHz maser exhibiting the strong inversion over a wide range of specific column densities. All maser transitions exhibit similar decay rates, between $1$ s$^{-1}$ and $12$ s$^{-1}$, where we note that decay rates increase with the number density and specific column density. 
%After that, we adopt conditions based on the shock modeling of \citet{hollenbach:13}, generally representative for water masers associated with a C- or J-shock.

The predicted anisotropic pumping parameter, $q_{\mathrm{anis}}$, is generally highest for the $183$ GHz transition, compared to the $22$ GHz and $321$ GHz transitions. It is lowest for those specific column densities where population inversion is maximal, and diverges when maser transitions start to thermalize. The $q_{\mathrm{anis}}$-parameter for the $22$ GHz and $321$ GHz transitions is below $0.1\ \%$, meaning that for a $\tau_{\mathrm{m}} \sim 15$ (unsaturated) maser, $\lesssim 2 \ \%$ polarization will be produced due to anisotropic pumping. For the $183$ GHz maser, higher $q_{\mathrm{anis}}$-parameters are expected, in particular for low(er) density gas. There, for masers with $\tau_{m}^{\perp} \gtrsim 1$, $q_{\mathrm{anis}}\lesssim 0.7 \ \%$, leading to polarization fractions of $\lesssim 10.5 \ \%$ for (unsaturated) masers with $\tau_{\mathrm{m}} \sim 15$. For the $183$ GHz transition excited at higher densities, $q_{\mathrm{anis}}$ is approximately halved compared to lower density masers, and flips sign to become positive when the maser starts to thermalize at high specific column densities. A sign flip in the $q_{\mathrm{anis}}$ parameter entails a $90^o$-flip in polarization direction of anisotropically pumped (unsaturated) masers. 

\begin{figure}[ht!]
\centering
    \includegraphics[width=0.4\textwidth]{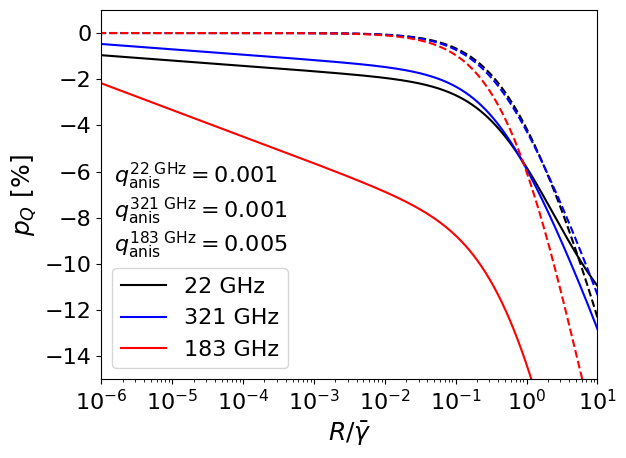}
    \caption{Predicted linear polarization fraction (definition: Eq.~\ref{eq:linpolfrac}) as a function of maser saturation level of H$_2$O maser transitions, with (solid lines) and without (dashed lines) the inclusion of anisotropic pumping parameters.} 
    \label{fig:water_prop}
\end{figure}

We now proceed to implement the results of the excitation analyses into a proper polarized radiative transfer simulation, using the CHAMP program package \citep{lankhaar:19}. In order to highlight the impact of anisotropic pumping on the polarization of the maser lines, we perform simulations with and without including anisotropic pumping. We adopted anisotropic pumping parameters $q_{\mathrm{anis}}=0.1\ \%$ for the $22$ GHz and $321$ GHz transitions, and $q_{\mathrm{anis}}=0.5\ \%$ for the $183$ GHz transition. Fig.~(\ref{fig:water_prop}) reports the predicted polarization fraction as a function of the maser saturation degree, $R/\bar{\gamma}$, where $R$ is the rate of stimulated emission. We note that due to the low degrees of anisotropic pumping, polarization degrees for the $22$ GHz and $321$ GHz masers are predicted to be low in unsaturated masers $R<\bar{\gamma}$. Additionally, when these masers saturate, anisotropic pumping barely affects the predicted levels of polarization generated by saturation polarization. In contrast, for the $183$ GHz maser, significant polarization, up to $7\ \%$ is predicted for unsaturated masers, while a significant boost in the predicted polarization fraction is predicted for saturated masers, compared to simulations that consider only saturation polarization alone.

%\begin{table}[ht!]
%    \caption{Summary of the LVG excitation analysis of water masers. }
%\label{tab:B_sensitivity}
%    \centering
%    \begin{tabular}{l c c}
%    \hline \hline
%    Technique & Sensitivity & Equation  \\
%    \hline
%Circular pol. & $B_{\mathrm{los}}$ & Eq.~(\ref{eq:pV}) \\ 
%Zeeman broad. & $B^2$, $B_{\mathrm{los}}^2$ & Eq.~(\ref{eq:zeeman_broad}) \\ 
%Linear pol. & $B_{x_\mathrm{pos}}^2 - B_{y_\mathrm{pos}}^2$, $B_{x_\mathrm{pos}}B_{y_\mathrm{pos}}$  & Eq.~(\ref{eq:pQ_fac}) \\ 
%\hline
%GK effect &  $\frac{B_{x_\mathrm{pos}}^2- B_{y_\mathrm{pos}}^2}{B^2}$, $\frac{B_{x_\mathrm{pos}}B_{y_\mathrm{pos}}}{B^2}$ & \\
%Dust polarization &  $\frac{B_{x_\mathrm{pos}}^2- B_{y_\mathrm{pos}}^2}{B^2}$, $\frac{B_{x_\mathrm{pos}}B_{y_\mathrm{pos}}}{B^2}$ & \\
%\hline
%    \end{tabular}
%        \raggedright \justify \vspace{-0.2cm}
%\textbf{Notes:} The magnetic field components are divided up into their line-of-sight component, $B_{\mathrm{los}}$ and their two components in the plane-of-the-sky, $B_{x_\mathrm{pos}}$ and $B_{y_\mathrm{pos}}$.
%
%\end{table}

\subsection{Class I CH$_3$OH masers}
Class I CH$_3$OH masers are thought to be excited in shocked gas toward high-mass star-forming regions, where they are often associated with an outflow structure. Class I CH$_3$OH masers can be roughly divided into three families of maser transitions:  $(J+1)_{-1} \to J_0$ E-type transitions, the $(J+1)_0 \to J_1$ A-type transitions, and $J_2 \to J_1$ E-type transitions, where the latter family of transitions occur around $25$ GHz. Class I CH$_3$OH masers of the $25$ GHz family are excited at densities $>10^6$ cm$^{-3}$, in contrast to the other class I CH$_3$OH masers that may be population inverted at densities as low as $10^4$ cm$^{-3}$. Still, in general, the strongest class I CH$_3$OH masers are expected at high densities of $10^7$ cm$^{-3}$ \citep{leurini:16}. We investigate the anisotropic pumping of at least one transition of each of the three families of class I CH$_3$OH masers. We investigate the $4_{-1} \to 3_0$ E-type transition that occurs at $36$ GHz, the $7_0 \to 6_1$ A-type transitions at $44$ GHz, and the $5_2 \to 5_1$ E-type transition at $25$ GHz. For all these transitions, circular polarization has been detected due to the Zeeman effect \citep{sarma:09, momjian:12b, momjian:17, sarma:20}. We include also the $8_0 \to 7_1$ A-type transition at $95$ GHz, because an observational analysis of its linear polarization properties in relation to the $44$ GHz transition has been performed previously \citep{kang:16}.
\begin{figure}[ht!]
\centering
    \begin{subfigure}[b]{0.4\textwidth}
      \includegraphics[width=\textwidth]{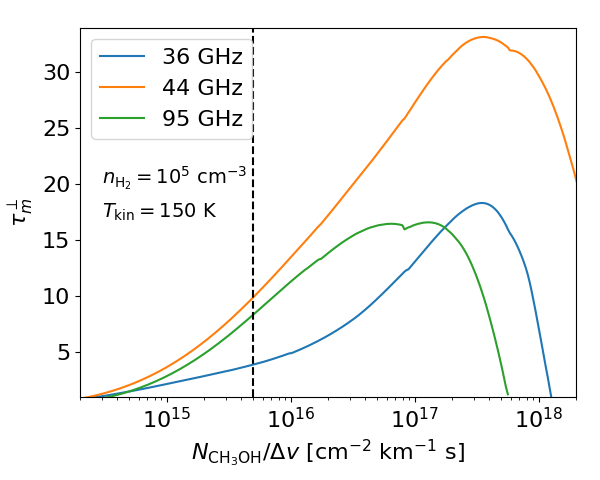}
      \caption{}
      \label{fig:methanol_5}
    \end{subfigure}
%\hfill
    \begin{subfigure}[b]{0.4\textwidth}
      \includegraphics[width=\textwidth]{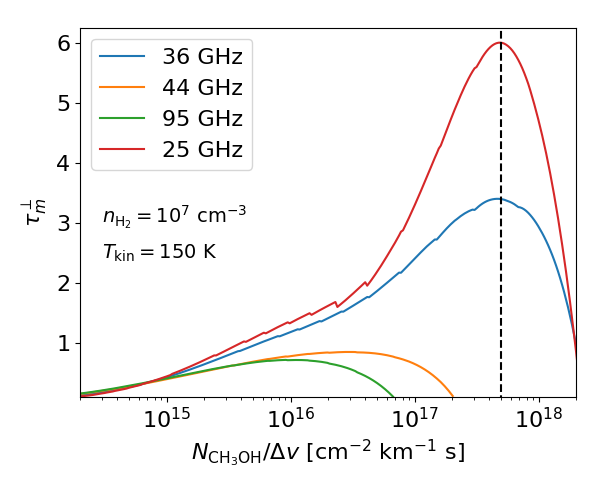}
      \caption{}
      \label{fig:methanol_7}
    \end{subfigure}
\caption{Plots of the maser optical depth of class I CH$_3$OH maser transitions as a function of the specific column density. Vertical dashed lines indicate the limit of physically reasonable maser extents. Excitation conditions are indicated inside the figure. } 
\label{fig:methanol_maser}
\end{figure}

\begin{figure*}[ht!]
\centering
    \begin{subfigure}[b]{0.4\textwidth}
      \includegraphics[width=\textwidth]{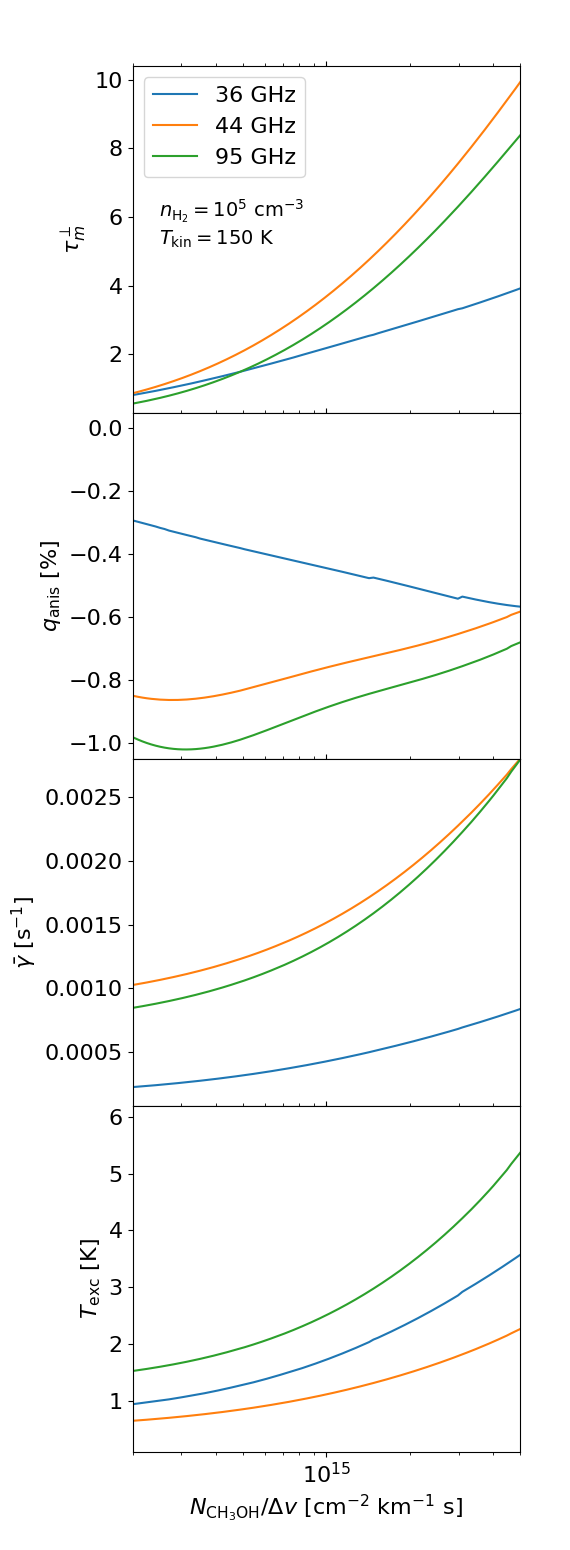}
      \caption{}
      \label{fig:methanol_5}
    \end{subfigure}
%\hfill
    \begin{subfigure}[b]{0.4\textwidth}
      \includegraphics[width=\textwidth]{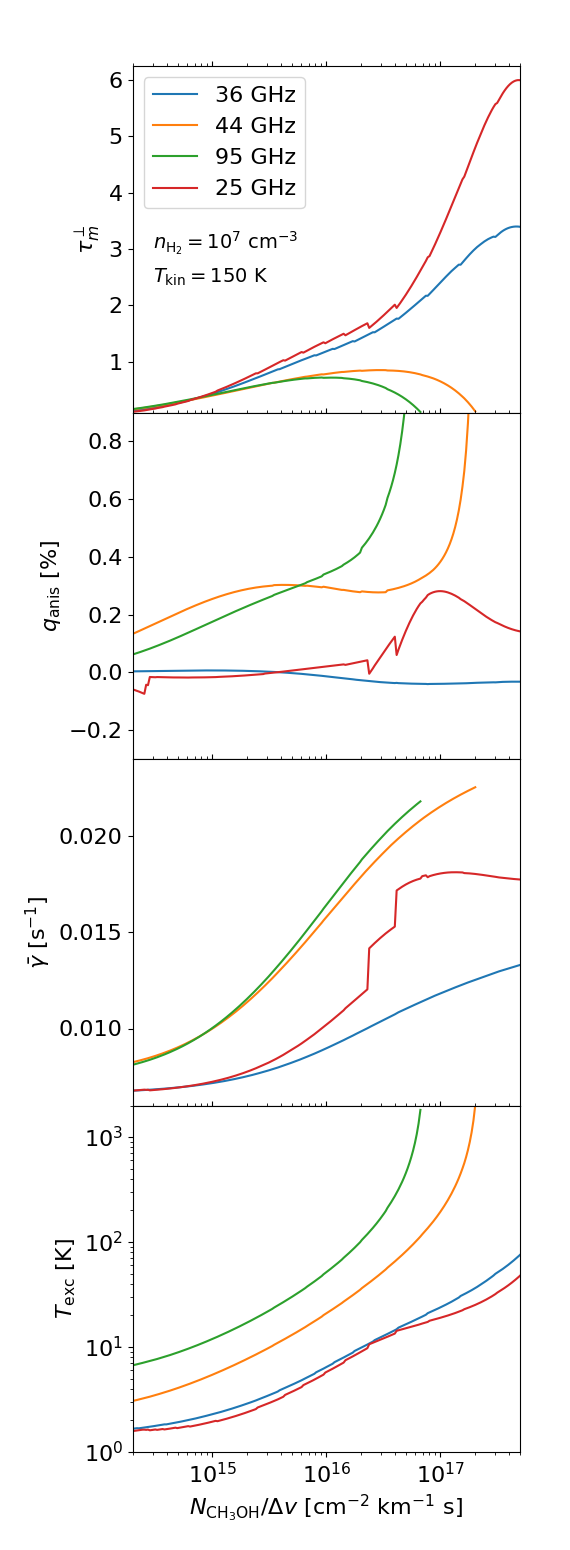}
      \caption{}
      \label{fig:methanol_7}
    \end{subfigure}
\caption{Plots of the maser (anisotropic) pumping parameters of class I CH$_3$OH maser transitions as a function of the specific column density. Excitation conditions are indicated inside the figure. Plots are given for the maser optical depth (upper row), anisotropic pumping factor (second row), decay rate (third row) and excitation temperature (lower row) of common maser transitions at $36$ GHz, $44$ GHz, $95$ GHz and $25$ GHz.} 
\label{fig:methanol_standard}
\end{figure*}

We modeled the excitation and anisotropic pumping of Class I CH$_3$OH masers. The $A$- and $E$-symmetry types of CH$_3$OH may be considered as chemically unconnected species and their excitation analysis is performed separately. We included the excitation of the $256$ lowest levels for both symmetry types, using collisional rate coefficients from \citet{rabli:10}. The maser region was approximated as disk-like LVG geometry, adopting a modest aspect ratio of $\epsilon=10$. We considered H$_2$ number densities $n_{\mathrm{H}_2}  = 10^5\ \mathrm{cm}^{-3}$ and $n_{\mathrm{H}_2}  = 10^7\ \mathrm{cm}^{-3}$, representative of sources that do not, and do exhibit the $25$ GHz masers. We furthermore assume a gas temperature of $T=150$ K and set the background radiation field at $T_{\mathrm{B}}=2.73$ K.

Results of the maser excitation modeling of class I CH$_3$OH masers are given in Fig.~(\ref{fig:methanol_maser}). We discuss the operation of class I CH$_3$OH masers, before we move on to discuss the anisotropic pumping parameters. Maser action is predicted for all the investigated transitions at high number densities $>10^6$ cm$^{-3}$, whereas for lower densities, the family of $25$ GHz transitions population inversion is suppressed, in agreement with the findings of \citet{leurini:16}. Similar to the behavior of H$_2$O maser transitions, the maser optical depth, $\tau_{\mathrm{m}}^{\perp}$, is a strong function of the specific column density, slowly rising until it reaches maximal inversion, where after a precipitous fall in population inversion occurs, and the levels are thermalized. Strongest maser action is expected at lower densities, but we should note that maximal population inversion occurs, for the $36$ GHz, $44$ GHz and $95$ GHz transitions, at specific column densities on the order of $N_{\mathrm{CH_3OH}}/\Delta v \sim 5\times 10^{17} \mathrm{cm}^{-2}\ \mathrm{km}^{-1} \ \mathrm{s}$, that for $n_{\mathrm{H}_2}=10^5 \ \mathrm{cm}^{-3}$ gas corresponds to maser sizes of $d \sim 5\times 10^{18} \ \mathrm{cm}$ ($3\times 10^5\ \mathrm{AU}$), assuming $x_{\mathrm{CH_3OH}} \sim 10^{-6}$. Such large masers cannot be reasonably expected, taking into account the requirement of velocity coherence. Indeed, also for the high density simulations, maximal population inversion (for the $25$ GHz and $36$ GHz masers) is expected at unphysically large masers of $\sim 5000$ AU. We indicate the limit of physically reasonable maser extents, which we take as $d \lesssim 1000$ AU, corresponding to $N_{\mathrm{CH_3OH}}/\Delta v \lesssim 5 \times 10^{15}\ \mathrm{cm}^{-2}\ \mathrm{km}^{-1} \ \mathrm{s}$ for $n_{\mathrm{H}_2}=10^5 \ \mathrm{cm}^{-3}$ and $N_{\mathrm{CH_3OH}}/\Delta v \lesssim 5\times 10^{17}\ \mathrm{cm}^{-2}\ \mathrm{km}^{-1} \ \mathrm{s}$ for $n_{\mathrm{H}_2}=10^7 \ \mathrm{cm}^{-3}$, inside Fig.~(\ref{fig:methanol_maser}). Indeed, while maximal population inversion occurs after these column densities, maser action is predicted.  % In the following, we discuss the anisotropic pumping properties of Class I methanol masers.

In Figs.~(\ref{fig:methanol_maser})-(\ref{fig:methanol_standard}), it may be noted that the property functions in some places show discontinuities. This is the result of the root-finding algorithm converging to a different, but nearby, solution in parameter space. The maser properties in general do not deviate by more than $\sim 10\%$ between discontinuities, so these features of the root-finding algorithm do not represent an issue for our analysis and we proceed to discuss the results of our excitation modeling.

We proceed to analyze the anisotropic pumping properties of class I CH$_3$OH masers excited in shock geometries with $d \lesssim 1000$ AU. The predicted anisotropic pumping parameters, $q_{\mathrm{anis}}$, are generally rather high for all the excited and investigated transitions at the low density of $n = 10^5$ cm$^{-3}$. Interestingly, at low density, the $q_{\mathrm{anis}}$ increases with the specific column density for the $36$ GHz transition, while it slightly, but consistently, drops for both the $95$ GHz and $44$ GHz transitions. In general, the $95$ GHz and $44$ GHz transitions exhibit similar behavior in all the investigated parameters, which can be ascribed to them belonging to the same family of maser transitions. At higher densities, $q_{\mathrm{anis}}$ is substantial, around $0.3\ \%$, but, contrary to lower densities, positive for the $95$ GHz and $44$ GHz transitions. Also, the anisotropy of the pumping increases with the specific column density. The $36$ GHz transition is only marginally anisotropically pumped at higher densities. The $25$ GHz transition is excited only at higher densities, where it exhibits significant anisotropic pumping when its population inversion is maximal. Adopting the decay rates that we modeled and assuming CMB as seeding radiation, all investigated transitions saturate around $\tau \sim 15$. Anisotropically pumped transitions, with $q_{\mathrm{anis}}\sim 0.3\ \%-1.0 \ \%$ will thus lead to linear polarization fractions up to $4.5\ \% - 15 \%$ for unsaturated masers. 
\begin{figure}[ht!]
\centering
    \begin{subfigure}[b]{0.4\textwidth}
      \includegraphics[width=\textwidth]{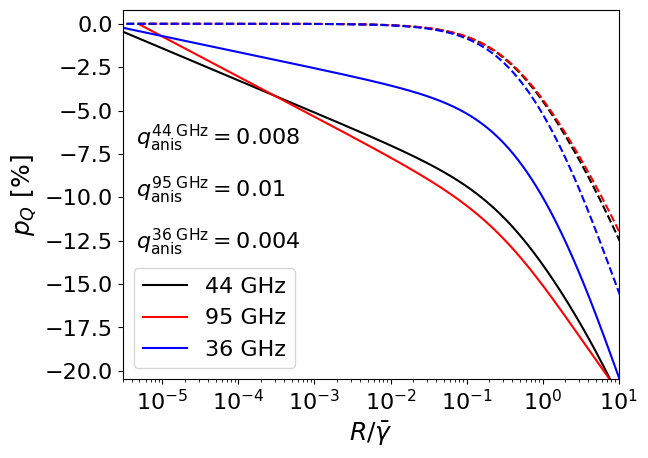}
      \caption{}
      \label{fig:prop_ten}
    \end{subfigure}
%\hfill
    \begin{subfigure}[b]{0.4\textwidth}
      \includegraphics[width=\textwidth]{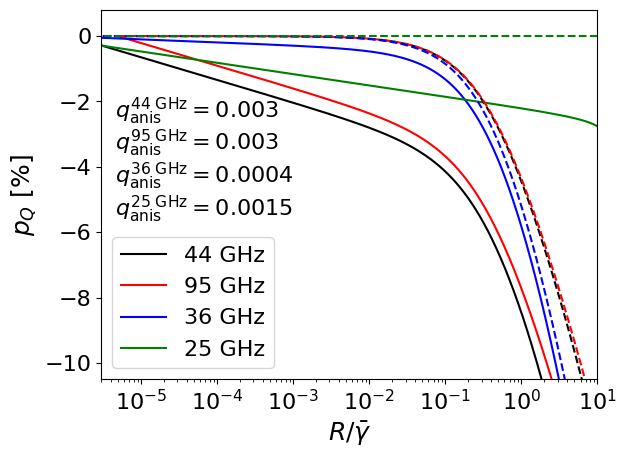}
      \caption{}
      \label{fig:prop_dense}
    \end{subfigure}
\caption{Predicted linear polarization fraction (definition: Eq.~\ref{eq:linpolfrac}) as a function of maser saturation level of class I CH$_3$OH maser transitions at low (a) and high (b) densities. Solid lines indicate simulations with anisotropic pumping, dashed lines indicate simulations without anisotropic pumping.} 
\label{fig:methanol_prop}
\end{figure}

In Fig.~(\ref{fig:methanol_prop}) we implement the results of the excitation analyses into proper polarized radiative transfer simulation. We perform simulations with and without including anisotropic pumping. Simulations were performed for class I CH$_3$OH masers excited in low and high density gas. We indicated the anisotropic pumping parameters that we adopted inside the figures. All the investigated maser transitions in low density gas are significantly affected by their anisotropic pumping. Both the $44$ GHz and $95$ GHz transitions produce polarization fractions exceeding $10\%$ as unsaturated masers, which impacts the saturation polarization at later stages of the propagation. A more modest, yet tangible, effect is also seen for the $36$ GHz maser excited in low density gas. In contrast, in high-density gas, the $36$ GHz transition is only marginally affected by its anisotropic pumping. Anisotropic pumping of the $44$ GHz and $95$ GHz CH$_3$OH masers is more modest, too. Anisotropic pumping in the $25$ GHz maser produces polarization fractions on the order of some percents.  %We note that due to the low degrees of anisotropic pumping, polarization degrees for the $22$ GHz and $321$ GHz masers are predicted to be low in unsaturated masers $R<\bar{\gamma}$, while for saturated masers, anisotropic pumping barely affects the predicted levels of polarization due to regular maser polarization due to saturation. In contrast, for the $183$ GHz maser, significant polarization, up to $7\ \%$ is predicted for unsaturated masers, while a significant boost in the predicted polarization fraction is predicted for saturated masers, compared to simulations that consider only saturation polarization alone.

%We considered H$_2$ number densities $\log_{10} (n_{\mathrm{H}_2} \ \mathrm{cm}^{-3} ) = 5, \ 6, \ 7, \mathrm{and} \ 8$, with $n_{\mathrm{H}_2} = 10^{7}\ \mathrm{cm}^{-3}$ as our fiducial case, and a range of CH$_3$OH specific column densities, $\frac{N_{\mathrm{CH_3OH}}}{\Delta v}$, between $10^{14} \ \mathrm{cm}^{-2}\ \mathrm{km}^{-1} \mathrm{s} $ and $10^{17} \ \mathrm{cm}^{-2}\ \mathrm{km}^{-1} \mathrm{s} $, with our fiducial case set to $10^{16} \mathrm{cm}^{-2}\ \mathrm{km}^{-1} \mathrm{s} $. We considered a range of temperatures, $T=50,\ 100,\ 150,$ and $200$ K, where we put $T=100$ K as our fiducial value.

\subsection{SiO masers}
While first discovered toward Orion KL \citep{snyder:74}, the most common SiO masers are excited close to evolved stars, in their extended atmosphere just before the dust sublimation zone \citep{gray:12}. The strongest SiO maser transitions occur in the first vibrationally excited state, where relaxation through vibrational de-excitation decreases in rate with $J$, when the de-excitation transitions are optically thick \citep{lockett:92}. Different excitation models for SiO masers have emphasized the importance of various features of SiO maser excitation, such as the influence of line-overlap \citep{olofsson:81, soria:04, desmurs:14}, nonlocal radiative transfer effects \citep{gonzalez:97, yun:12}, or their variability in relation to the variability of the host star \citep{humphreys:02}. SiO masers are often highly polarized \citep{kemball:97}, which is commonly ascribed to their anisotropic pumping that is a result of directional radiation from the central star \citep{western:83c}.
\begin{figure*}[ht!]
\centering
    \begin{subfigure}[b]{0.4\textwidth}
      \includegraphics[width=\textwidth]{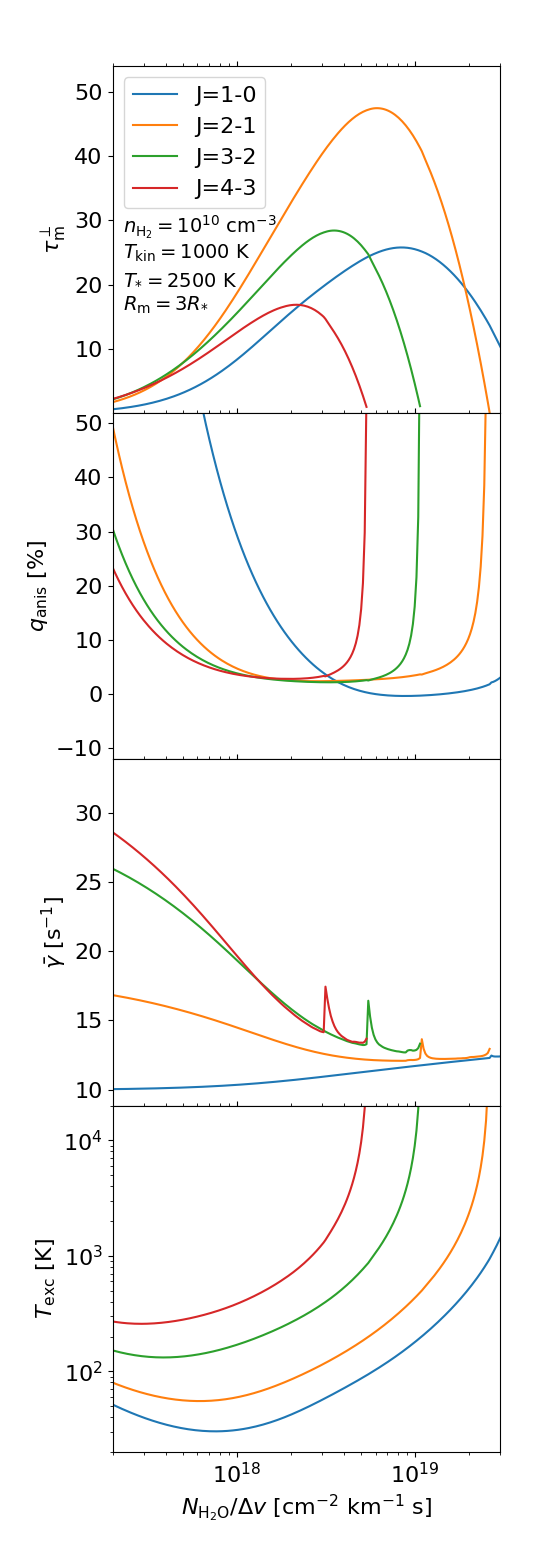}
      \caption{}
      \label{fig:sio_low}
    \end{subfigure}
%\hfill
    \begin{subfigure}[b]{0.4\textwidth}
      \includegraphics[width=\textwidth]{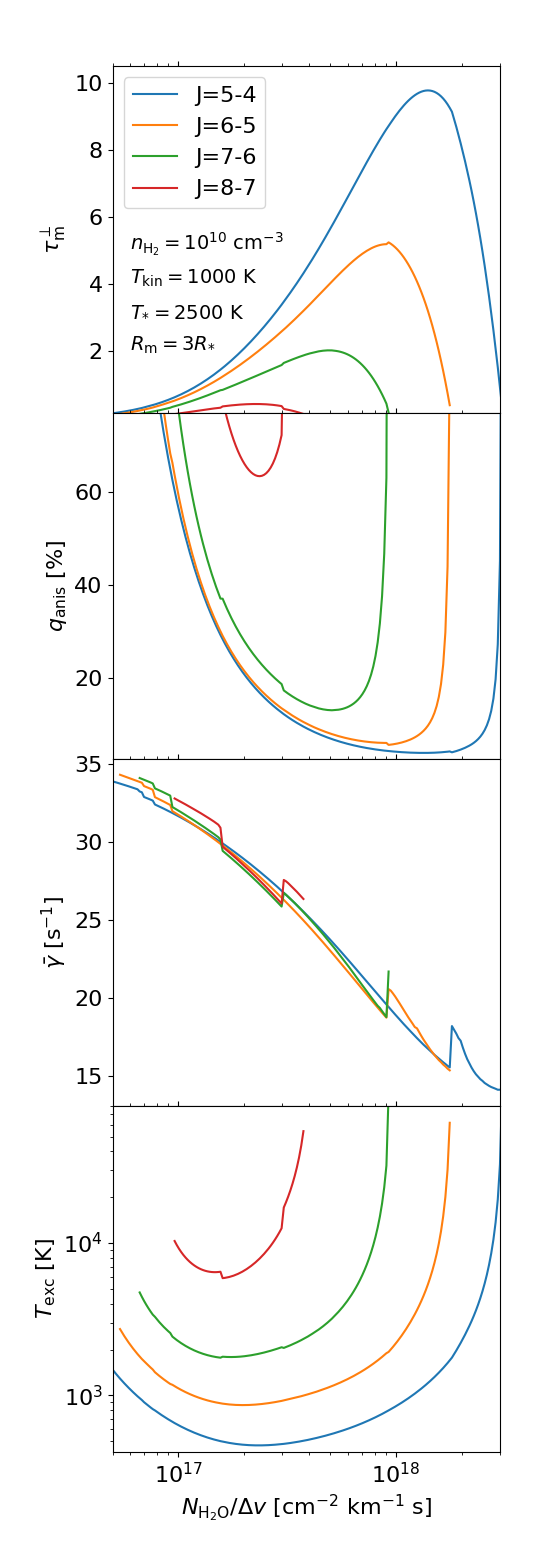}
      \caption{}
      \label{fig:sio_high}
    \end{subfigure}
\caption{Plots of the maser (anisotropic) pumping parameters of SiO maser transitions as a function of the specific column density. Excitation conditions are indicated inside the figure. Plots are given for the maser optical depth (upper row), anisotropic pumping factor (second row), decay rate (third row) and excitation temperature (lower row) of common maser transitions $J \to J-1$ ($J=1,\cdots,8$), in the first vibrationally excited state.} 
\label{fig:sio_standard}
\end{figure*}

We modeled the excitation and (anisotropic) pumping of SiO masers in the first vibrational state toward an oxygen-rich AGB star. We consider an excitation analysis adopting a number density and temperature of $n_{\mathrm{H}_2}=10^{10}\ \mathrm{cm}^{-3}$ and $T=1000$ K, that is representative of SiO maser clumps with an enhanced density \citep{lockett:92}. We modeled the maser region in a disk-like geometry, where we assume an aspect ratio $\epsilon=2$. The excitation analysis was performed including the first two vibrationally excited states, including rotational levels up to $J=40$ in each vibrational state. Line-overlap with other molecules was not taken into account. The vibrationally resolved collisional transition rates were taken from \citet{yang:18}. In addition to the molecular excitation conditions, it is of importance to properly model the radiation environment of the SiO masers. In particular, the radiation field of the central star affects the vibrational transitions significantly \citep{lockett:92}, but also impacts the anisotropic pumping of the maser. In our excitation modeling, we included the radiation field of a star, having a black-body radiation field of $T_{*}=2500$ K. We assumed the SiO maser clumps are situated at $3R_{*}$.

As for the CH$_3$OH masers, it may be noted from Fig.~(\ref{fig:sio_standard}) that the property functions at places show discontinuities. Again, since the maser properties in general do not deviate by more than $\sim 10\%$ between discontinuities, these features do not represent an issue for our analysis and we proceed to discuss the results of our excitation modeling.

We predicted maser action in the first $7$ rotational transitions of the first vibrational state. Maser action is predicted to be strongest for the $J=2 \to 1$ transition for all but the highest column densities, which is not in line with observations, but a common feature of SiO maser excitation modeling \citep{lockett:92}. After the $J=2 \to 1$ transition, maser optical depths are predicted to gradually lower, as well as the specific density at which maximal maser action occurs. By including collisional and radiative, alike, we predict decay rates for the $ J\to J-1$ transitions, roughly adhering to $\bar{\gamma} / \mathrm{s}^{-1} \approx 10 + 2J$, which is higher than the commonly assumed $5v \ \mathrm{s}^{-1}$, where $v$ is the vibrational quantum number, that follows from only taking into account radiative decay \citep{elitzur:92}. The predicted excitation temperatures roughly adhere to $T_{\mathrm{exc}}\approx 30 \times 2^J\ \mathrm{K}$. From the excitation temperatures and decay rates we may derive saturation optical depths, roughly adhering to $\tau_{\mathrm{sat}} \approx 17-J$.

Anisotropic pumping for the vibrationally excited SiO masers is predicted to be high, in particular for the high $J$ transitions (see Fig.~\ref{fig:sio_high}). For transitions $J\to J-1$ up to $J=6$, we predict anisotropic pumping parameters of $q_{\mathrm{anis}} \sim 5\ \%$ for the specific column densities where significant maser action is predicted. The predicted anisotropic pumping parameters of higher $J$ transitions precipitously rise to $q_{\mathrm{anis}} \sim 18\ \%$ for the $J=7-6$ transition and $q_{\mathrm{anis}} \sim 65\ \%$ for the $J=8-7$ transition. Strong anisotropic pumping in the higher $J$ transitions is likely due to the precipitous rise in excitation temperature \citep[for more discussion between the relation between relative population inversion and anisotropic pumping polarization yields, see][]{nedoluha:90}. Adopting the saturation optical depths discussed earlier, the predicted $q_{\mathrm{anis}}$ parameters for the lower $J$ transitions (up to $J=6\to 5$), yield linear polarization fractions of $50-75\%$ when the maser saturates, and up to $100\%$ for the higher $J$ transitions. Polarization due to maser saturation can either enhance or diminish the polarization due to anisotropic pumping, depending on the orientation of the magnetic field with respect to the propagation direction.

\begin{figure}[ht!]
\centering
    \begin{subfigure}[b]{0.4\textwidth}
      \includegraphics[width=\textwidth]{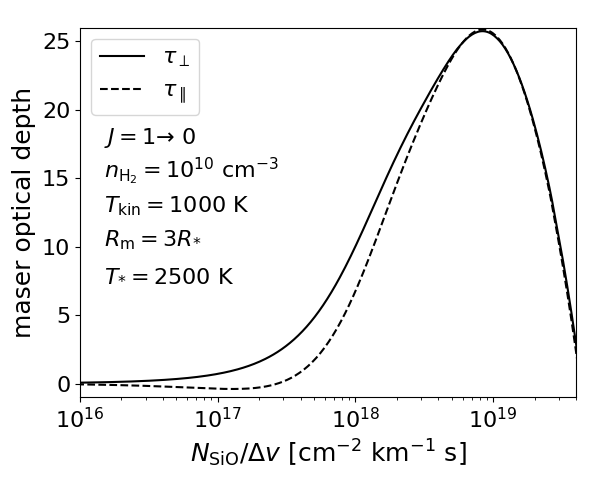}
      \caption{}
      \label{fig:sio_tau_low}
    \end{subfigure}
%\hfill
    \begin{subfigure}[b]{0.4\textwidth}
      \includegraphics[width=\textwidth]{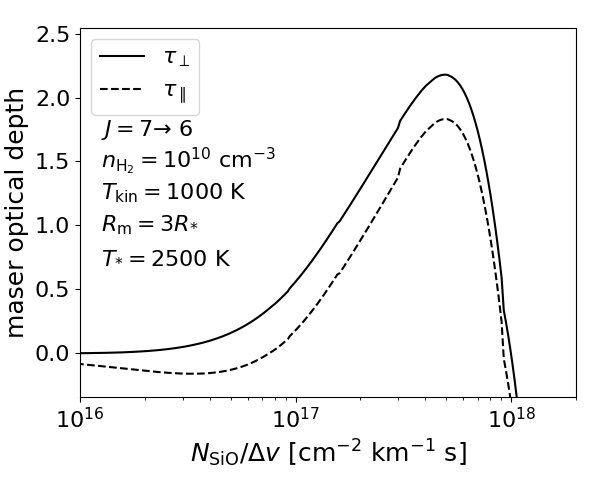}
      \caption{}
      \label{fig:sio_tau_high}
    \end{subfigure}
\caption{Predicted optical depths, parallel and perpendicular to the magnetic field, as a function of maser saturation level for the SiO (a) $J=1\to 0$ and (b) $J=7\to 6$ transitions in the first vibrationally excited state.} 
\label{fig:sio_tau}
\end{figure}

For the highly anisotropically pumped vibrationally excited SiO masers, it is interesting to directly compare the optical depths parallel and perpendicular to the magnetic field, as a complement to the proxy parameter $q_{\mathrm{anis}}$. In Fig.~(\ref{fig:sio_tau}), we plot the optical depth of both linearly polarized components of the radiation field, for the $J=1\to0 $ and $J=7\to 6$ transitions in the first vibrationally excited state. There, it is interesting to note, that in particular for low specific optical depths, the perpendicular component of the (maser) optical depth is positive, while the parallel component is negative. In practice, this means that under such conditions, the radiation field component that is polarized perpendicular to the magnetic field will be amplified, while the component that is polarized parallel to the magnetic field will be in absorption. Radiation that is processed by such a maser will be fully polarized. This phenomenon is a manifestation of a dichroic maser, that was predicted by \citet{ramos:05}, using simplified, but polarization resolved, excitation modeling of SiO masers. We note, however, that dichroic masing occurs for weak masers, at low specific column densities. In stronger masers both of the polarization components of the radiation field will be amplified. For these masers, the difference in optical depth between the polarization components will lead to the partial polarization of the radiation.

%If we assume the SiO masers to be excited in an outflow with velocity, $v$, at radius $R$, then the relevant specific column density is,
%\begin{align}
%\frac{N_{\mathrm{SiO}}}{\Delta v} &= 7.48 \times 10^{17} \ \mathrm{cm}^{-2}\ \mathrm{km^{-1}} \mathrm{s} \left( \frac{n_{\mathrm{H}}_2 x_{\mathrm{SiO}}}{10^5 \ \mathrm{cm}^{-3}} \right)\nonumber \\ &\times \left(\frac{v}{10 \ \mathrm{km/s}} \right)^{-1} \left(\frac{R}{5 \ \mathrm{AU}} \right),
%\end{align}
%with a certain aspect ratio.

%We model this by modeling the radiative transfer through the circumstellar envelope toward the resonant frequencies of the SiO transitions that we consider in the excitation analysis. We model the radiative transfer assuming it comes exclusively from dust emission or the central star. We modeled the dust radiative transfer using the opacity coefficients of \citet{ossenkopf:...} and assuming a density profile of $...$ and temperature profile as $...$, that we base the stellar and circumstellar wind parameters on TX Cam \citep{debeck:..., willens:...}. We plot the $\mu$-dependent background radiation field in units of photon occupation number for the $v=1-0,\ J=1-0$ transition and the $v=1,\ J=1-0$ transition in Fig.~(\ref{fig:SiO_rad}). 

%\dot{M} = 6.5e-6
%Teff = 2779 K
%R_star = 460 solar radii
%R_maser = 4*R_star = 1840 solar radii = 
%v_exp = 18 km/s
%---> n_H = 1.5e11 cm-3 * 6.5e-2 / (1.28^2 * 18) = 3.3e8 cm-3
%

\section{Discussion}
\subsection{Anisotropic pumping}
The maser species that exhibit significant anisotropic pumping can be roughly divided into two classes: (i) maser species that are excited in the vicinity of a strong and directional IR source, and (ii) maser species that are excited in an anisotropic geometry, such as a shock. These two classes were represented in our simulations by (i) vibrationally excited SiO masers, and (ii) H$_2$O and class I CH$_3$OH masers. In general, we found that the influence of a strong and directional IR source, lead to strongly anisotropically pumped masers, that may produce up to $100\%$ polarization. Masers that are excited in an anisotropic geometry generally show lower degrees anisotropic pumping, and yield polarization fractions of $\lesssim 15\%$ due to anisotropic pumping. For both classes of masers, anisotropic pumping is dependent on the excitation conditions, and highly dependent on the transition.

Anisotropic pumping due to an external IR source has been recognized as a way to generate the high polarization yields of, in particular, vibrationally excited SiO masers \citep{western:83c,kemball:09}. Our simulations are the first to attach quantitative estimates of the degree of anisotropic pumping, where we confirm that indeed polarization degrees up to $100\%$ may be explained by anisotropic pumping alone. 

From the clear association between anisotropic pumping and a strong IR source, it is interesting to note that recent works have observed a causal relation between an accretion burst, associated with a flare in IR luminosity, and a flare in the maser luminosity, of in particular the $6.7$ GHz class II CH$_3$OH maser \citep{hunter:18, szymczak:18, moscadelli:17, burns:20, stecklum:21}, but also H$_2$O and OH masers \citep{macleod:18}. We predict that these flaring masers should be associated with an increase in their linear polarization fraction, as the directional IR radiation source is an important feature of the excitation of these maser sources. Polarization observations of maser sources associated with accretion bursts can then yield important information on the magnetic field and its relation to accretion events. %While we have not modeled class II methanol masers or water masers in the vicinity of an IR source, we have been able to indicate a clear association between anisotropic pumping and a strong IR source, from the modeling of SiO masers. 

The anisotropic pumping of masers excited in an anisotropic geometry, such as a shock, has hitherto not been recognized as an efficient polarization mechanism of masers. Indeed, while it will likely not produce linear polarization fractions exceeding $20\%$, masers excited in an anisotropic geometry may be significantly and detectably polarized due to anisotropic pumping. In sections \ref{sec:galwater} and \ref{sec:methanol}, we discuss the implications of anisotropic pumping for the polarization of galactic and extragalactic H$_2$O masers, and class I CH$_3$OH masers, respectively.

We performed simulations of axisymmetric systems, where we assumed that the symmetry axis is aligned with the magnetic field, and perpendicular to line of sight. To generalize to arbitrary orientations of the magnetic field and the line of sight, we define the angle between the magnetic field and the symmetry axis, $\beta$, and the angle between the magnetic field and the line of sight, $\vartheta$. The polarization fraction due to anisotropic pumping is then
\begin{align}
p_Q^{\mathrm{anis}}(\beta,\vartheta) \simeq \frac{3\cos^2\beta -1}{2} \sin^2 \vartheta \ p_Q^{\mathrm{anis}},
\end{align}
where $p_Q^{\mathrm{anis}}$ is the polarization fraction in a system where $\beta=0$ and $\vartheta=\pi/2$. Note here, that the factor, $\frac{3\cos^2\beta -1}{2}$, which is the second Legendre polynomial of $\cos \beta$, can assume both positive and negative values. Thus, recalling the definition of $p_Q = (I_{\parallel} - I_{\perp}) / (I_{\parallel} + I_{\perp})$, with a varying $\beta$ in a source of equal $p_Q^{\mathrm{anis}}$, the polarization direction may flip 90$^o$.

\subsection{Anisotropic pumping and saturated masers}
The most common mechanisms that lead to the partial polarization of masers are through (i) maser saturation, where the beamed and strong maser radiation affects the population and alignment of the maser transition quantum states, and (ii) anisotropic pumping, where the maser transition quantum states are partially aligned from the outset, due to anisotropy in the pumping of the maser. Anisotropic pumping can produce polarization in unsaturated masers, while saturation polarization only is present in saturated masers. 

In the idealized case, for a $J=1\to 0$ transition, where no anisotropy in the pumping is assumed, and the maser is highly saturated $R \gg \bar{\gamma}$, while still magnetically aligned $g\Omega \gg R$, then it can be  shown analytically, that the polarization due to maser saturation converges to \citep{goldreich:73, lankhaar:21}
\begin{align}
\label{eq:gkk}
p_Q^{\mathrm{GKK}} = \begin{cases}
\frac{2-3\sin^2 \vartheta}{3\sin^2 \vartheta} \quad &\mathrm{for} \sin^2 \vartheta \geq \frac{1}{3} \\
1 \quad &\mathrm{for} \sin^2 \vartheta < \frac{1}{3}.
\end{cases}
\end{align}
In the following discussion, we take this equation as guidance, but we should note numerical simulations have indicated that the convergence to these levels of polarization are reached only at unphysically high levels of saturation \citep{nedoluha:90, lankhaar:19}. In addition, the analytical solution can not be formally extended to arbitrary angular momentum transitions. Numerical simulations find that with increasing angular momentum, the predicted degree of polarization due to maser saturation drops \citep{nedoluha:90, lankhaar:19}.

\begin{figure}[ht!]
\centering
    \begin{subfigure}[b]{0.4\textwidth}
      \includegraphics[width=\textwidth]{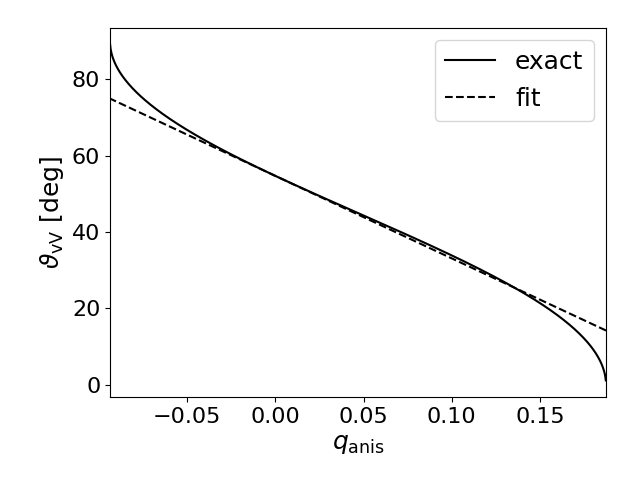}
      \caption{}
      \label{fig:vanVleck}
    \end{subfigure}

    \begin{subfigure}[b]{0.4\textwidth}
      \includegraphics[width=\textwidth]{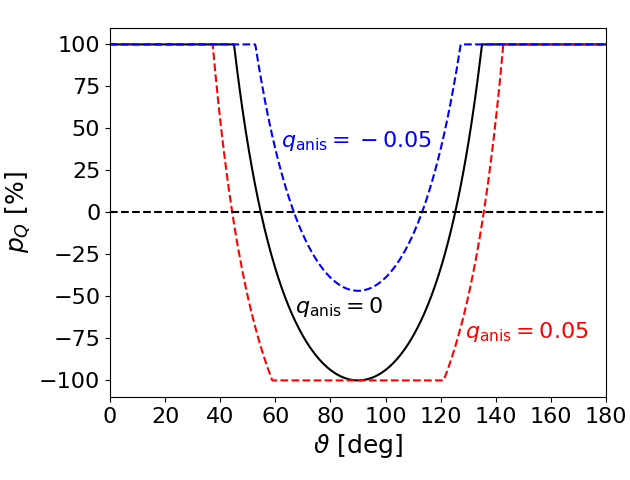}
      \caption{}
      \label{fig:gkk_anis}
    \end{subfigure}
%\hfill
\caption{Polarization properties of an anisotropically pumped saturated maser in the limit considered by \citet{goldreich:73} (see text preceding Eq.~\ref{eq:gkk}). Polarization properties are summarized through, (a) the van Vleck angle as a function of the anisotropic pumping degree, and (b) the polarization fraction as a function of the propagation angle for different degrees of anisotropically pumping.} 
\label{fig:gkk}
\end{figure}

One interesting property of the analytical solution represented in Eq.~(\ref{eq:gkk}), is that it predicts a sign change in the polarization fraction. A sign change in the polarization fraction occurs at the so-called van Vleck angle, $\vartheta_{\mathrm{vV}}$, which from Eq.~(\ref{eq:gkk}) coincides with the ``magic'' angle $\vartheta_{\mathrm{m}}\approx 54.7^o$. A sign change in the polarization fraction entails a $90^o$ flip in the polarization vector on the plane of the sky. A phenomenon that has been observed across both H$_2$O and SiO maser clumps \citep{vlemmings:06,tobin:19}. The sign change in the polarization fraction around the magic angle is a property that is reproduced in numerical simulations with $R\ll g\Omega$, but may be affected when $R\gtrsim g\Omega$ \citep{lankhaar:19}. When the maser is anisotropically pumped, we still predict a sign change. However, the propagation angle at which the sign change occurs, at the van Vleck angle, will not coincide with the magic angle: $\vartheta_{\mathrm{vV}}\neq \vartheta_{\mathrm{m}}$. In Fig.~(\ref{fig:vanVleck}), we plot the predicted van Vleck angle as a function of the anisotropic pumping degree. For small anisotropic pumping degrees, $|q_{\mathrm{anis}}|\lesssim 0.05$, the predicted van Vleck angle can be approximated by the function, 
\[
\vartheta_{\mathrm{vV}} \simeq \vartheta_{\mathrm{m}} - 2.16^o \left[\frac{q_{\mathrm{anis}}}{0.01} \right],
\]
indicating that for modestly anisotropically pumped masers, such as H$_2$O and CH$_3$OH masers, $\vartheta_{\mathrm{vV}} \approx \vartheta_{\mathrm{m}}$. For strongly anisotropically pumped masers, the van Vleck angle may deviate significantly from the magic angle. In extreme cases, for $q_{\mathrm{anis}}>3/16$ ($q_{\mathrm{anis}}<-3/32$) only perpendicular (parallel) polarization directions, with respect to the magnetic field direction, are predicted. This feature of anisotropically pumped masers has also been discussed by \citet{western:84} and \citet{elitzur:96}. 

The variance of the van Vleck angle with the degree of anisotropic pumping is representative of the interplay between anisotropic pumping and saturation polarization. In terms of polarization fraction yields, as can be seen in Fig.~(\ref{fig:gkk_anis}), anisotropic pumping can either boost or diminish the polarization produced through maser saturation. As we have seen, the polarization due to anisotropic pumping (in the unsaturated regime), $\sim -\tau_{\nu}^{(0)} q_{\mathrm{anis}}$, can either be parallel or perpendicular to the magnetic field direction, for negative and positive $q_{\mathrm{anis}}$, respectively. Thus, for masers that propagate with $\vartheta < \vartheta_{\mathrm{m}}$ and negative $q_{\mathrm{anis}}$, maser saturation and anisotropic pumping will be cooperative, while for $\vartheta < \vartheta_{\mathrm{m}}$ and positive $q_{\mathrm{anis}}$, they will be antagonistic.

\subsection{Production of circular polarization from anisotropically pumped masers}
\label{sec:lin_circ}
We considered the transfer of polarized radiation through an anisotropically pumped maser, where we assumed weak Zeeman splitting. Up to now, we considered a uniform magnetic field, but in this discussion, we relax this assumption and consider the consequences.

Because we assumed a uniform magnetic field, we have been able to decompose the radiation field in its polarization components parallel and perpendicular to the projected magnetic field direction on the plane of the sky. In terms of Stokes parameters: we only considered the Stokes $I$ and $Q$ parameters, where $I_{\parallel,\perp}=(I \pm Q)/2$. However, when the magnetic field changes direction over the propagation, then the symmetry axis of the maser molecules will not necessarily align with the symmetry axis of the polarization decomposition, and thus the third, Stokes $U$, parameter is required to fully describe the linear polarization direction of the radiation \citep{chandrasekhar:13}. In addition, when considering the transfer of radiation involving the Stokes $U$ parameter, we have to include the possible conversion of Stokes $U$ to Stokes $V$; a transformation that is related to the conversion of Stokes $I$ to Stokes $Q$ through the Kramers-Kronig relations \citep{kylafis:83, deguchi:85, wiebe:98}.

\begin{figure}[ht!]
\centering
    \includegraphics[width=0.45\textwidth]{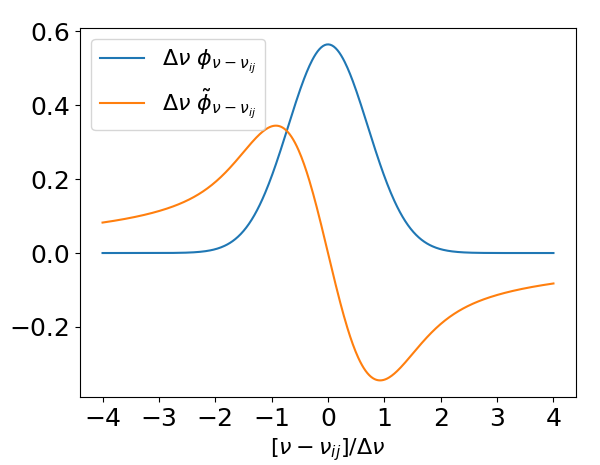}
    \caption{Normalized line profiles associated with the $\kappa_Q$ and $\tilde{\kappa}_Q$ transformation coefficients, assuming dominant Doppler broadening.} 
        \label{fig:kramers_profiles}
\end{figure}

We note the full polarized radiative transfer equation, for an unsaturated maser, that is fully aligned to the magnetic field, in an anisotropically pumped medium, exhibiting a negligible Zeeman effect \citep{lankhaar:19},
\begin{subequations}
\label{eq:full_pol_prop}
\begin{align}
\frac{d}{ds_{\Omega}}\begin{pmatrix} I \\ Q \\ U\\ V \end{pmatrix} = - 
\begin{pmatrix} 
\kappa_{I} & \kappa_{Q} & 0 & 0 \\
\kappa_{Q} & \kappa_{I} & 0 & 0 \\
0 & 0 & \kappa_{I} & \tilde{\kappa}_{Q} \\
0 & 0 & -\tilde{\kappa}_{Q} & \kappa_{I} 
\end{pmatrix}
+ 
\begin{pmatrix} \epsilon_{I} \\ \epsilon_{Q} \\ 0 \\ 0 \end{pmatrix},
\end{align}
where we suppressed the dependence of the Stokes parameters on the direction and frequency, $I = I(\nu,\Omega)$, for notational simplicity, and
\begin{align}
\kappa_I &= \left[k_0 + k_2 \frac{3\mu^2-1}{2\sqrt{2}} \right] \phi_{\nu-\nu_{ij}}, \\
\kappa_Q &= \left[k_2 \frac{3(1-\mu^2)}{2\sqrt{2}}\right] \phi_{\nu-\nu_{ij}}, \\
\epsilon_I &= \left[e_0 + e_2 \frac{3\mu^2-1}{2\sqrt{2}}\right] \phi_{\nu-\nu_{ij}}, \\
\epsilon_Q &= \left[e_2 \frac{3(1-\mu^2)}{2\sqrt{2}}\right] \phi_{\nu-\nu_{ij}},
\end{align}
are the usual polarized propagation coefficients, with the $k$- and $e$-parameters defined in Eqs.~(\ref{eq:rad_trans}), and where
\begin{align}
\tilde{\kappa}_Q = \left[k_2 \frac{3(1-\mu^2)}{2\sqrt{2}}\right]\tilde{\phi}_{\nu-\nu_{ij}},
\end{align}
\end{subequations}
is related to the $\kappa_Q$ propagation coefficient through the Kramers-Kronig relations, having a different line profile, which for a Doppler broadened, with line breadth $\Delta \nu$, line is $\tilde{\phi}_{\nu-\nu_{ij}} = \phi_{\nu-\nu_{ij}} \mathrm{erfi}([\nu-\nu_{ij}]/\Delta \nu)$. In Fig.~(\ref{fig:kramers_profiles}) we plot both profiles.

We may note Eqs.~(\ref{eq:full_pol_prop}) in an adjusted basis, that is found through matrix diagonalization. In the adjusted basis, the matrix propagation equation decomposes into four independent propagation equations. The first two are given in Eqs.~(\ref{eq:rad_trans}), while the other propagation equations note
\begin{subequations}
\begin{align}
\frac{d}{ds_{\Omega}} (iU \pm V) = -(\kappa_I \pm i \tilde{\kappa_Q}) (iU \pm V),
\end{align}
having the trivial solution
\begin{align}
(iU \pm V) = e^{-[\kappa_I \pm i \tilde{\kappa_Q}]s_{\Omega}} (iU \pm V)_0,
\end{align}
\end{subequations}
where $(iU \pm V)_0$ is the incoming radiation field. 

Suppose we have an incoming radiation field, described by a Stokes $U$ component of $U_0$, and a total intensity $I_0$. After propagating for a distance, $s$, the resulting yields for the Stokes $U$ and $V$ parameters are,
\begin{subequations}
\begin{align}
U &= e^{-\kappa_I s} \cos (\tilde{\kappa}_Q s) U_0,\\
V &= e^{-\kappa_I s} \sin (\tilde{\kappa}_Q s) U_0.
\end{align}
We can express the Stokes parameters in terms of the polarization fractions, by normalizing them with the total intensity: $I = I_0 e^{\kappa_I s}$,
\begin{align}
\frac{U}{I} &= \cos (\tilde{\kappa}_Q s) \frac{U_0}{I_0},\\
\frac{V}{I} &= \sin (\tilde{\kappa}_Q s) \frac{U_0}{I_0}.
\end{align}
\end{subequations}
from which it becomes clear that the transformation coefficient $\tilde{\kappa}_Q$, in effect rotates incoming Stokes $U$ radiation to Stokes $V$, and vice-versa. 

Note that the quantity $\tilde{\kappa}_Q s = -q_{\mathrm{anis}}\tau_0^{\mathrm{m}} \tilde{\phi}_{\nu-\nu_{ij}}$ is, excluding the vibrationally excited SiO masers, fairly small, $\lesssim 10\%$, for the investigated maser transitions. Therefore, in most cases, the yield of circular polarization from an incoming linearly polarized (Stokes $U$) radiation field, can be approximated by
\begin{align}
\frac{V}{I} \simeq -q_{\mathrm{anis}}\tau_{\nu}^{(0)} \tilde{\phi}_{\nu-\nu_{ij}} U_0.
\end{align}
We note that the spectral profile, captured in $\tilde{\phi}_{\nu-\nu_{ij}}$, is very similar to the S-shaped profile that is expected in the circular polarization profile of a Zeeman splitted line, thus the non-Zeeman circular polarization due to anisotropic pumping in combination with a changing magnetic field direction, may easily be mistaken for circular polarization due to the Zeeman effect. To estimate the impact of non-Zeeman circular polarization on magnetic field determination through Zeeman effects, we compare the circular polarization fraction at the maxima due to non-Zeeman effects,
\begin{subequations}
\begin{align}
p_V^{\mathrm{nZ}} \approx 0.61 \%   \left[\frac{-q_{\mathrm{anis}} \tau_{\nu}^{(0)}}{10 \%} \right]\left[\frac{U_0/I_0}{10 \%}\right],
\end{align}
to the circular polarization fraction at the maxima due to Zeeman effects \citep{lankhaar:23}
\begin{align}
%maybe better in terms of 
\label{eq:pV_zee}
p_V^{\mathrm{Z}} &\simeq 0.43 \%  \left[\frac{z}{2\ \mathrm{Hz\ \mathrm{mG}^{-1}}}\right] \left[\frac{\Delta v_{\mathrm{FWHM}}}{1\ \mathrm{km\ s^{-1}}}\right]^{-1} \nonumber \\ 
&\times \left[\frac{ \nu_{0}}{10\ \mathrm{GHz}} \right]^{-1} \left[\frac{B_{\mathrm{los}}}{100 \ \mathrm{mG}} \right],
\end{align}
\end{subequations}
where $\Delta v_{\mathrm{FWHM}}$ is the line width at half maximum, $z$ is the Zeeman coefficient, that is normalized to a representative value for non-paramagnetic species, $\nu_0$ is the line frequency and $B_{\mathrm{los}}$ is the magnetic field strength along the line of sight. 

Thus, for masers that are strongly linearly polarized ($\gtrsim 10\%$) due to anisotropic pumping, significant non-Zeeman circular polarization may be produced in a source where the magnetic field changes direction over the path of propagation. The production of circular polarization is quadratically related to the total linear polarization fraction, 
\begin{align}
p_l = \frac{\sqrt{Q^2 + U^2}}{I},
\end{align}
since both $U_0/I_0 \sim p_l$ and $q_{\mathrm{anis}}\tau_0^{\mathrm{m}}\sim p_l$ have a linear relation to the degree of linear polarization: $p_V^{\mathrm{nZ}} \sim p_l^2$. Therefore, non-Zeeman polarization is only appreciable for the strongly (linearly) polarized masers. Indeed, in simulations that compute the impact of a changing magnetic field on the production of non-Zeeman circular polarization, \citet{wiebe:98} found $p_V \sim p_l^2/4$, due to saturation polarization. We confirm here a similar relation for non-Zeeman circular polarization that is produced from linear polarization produced due to anisotropic pumping that is operative also for unsatured masers.

Finally, it should be noted that the mechanism commonly referred to as ``intensity dependent circular polarization'' \citep{nedoluha:94, lankhaar:19} is in fact a mechanism very similar to the mechanism illustrated above. The prime difference, is that it is not the magnetic field that rotates along the line of sight, but the symmetry axis of the maser molecules, that gets gradually realigned to the propagation direction when the maser radiation intensifies to match and exceed the magnetic precession rate: $R \sim g\Omega$. Indeed, also for ``intensity dependent circular polarization'', linear polarization is converted to circular polarization through the $\tilde{\kappa}_Q$ term in the propagation, and, in general, the relation $p_c \lesssim p_l^2$, to estimate the non-Zeeman circular polarization, may be expected here, too.

\subsection{Polarization of H$_2$O masers}
\label{sec:galwater}
The polarization of H$_2$O masers has been observed toward a variety of sources. H$_2$O masers show both circular and linear polarization toward high-mass star-forming regions \citep[e.g.,~][]{surcis:23}, but also evolved stars \citep{vlemmings:05b, vlemmings:06}. H$_2$O masers toward evolved stars are excited either in a shell-like structure around the central late-type star \citep{richards:12}, or in association with a fast and collimated outflow \citep[e.g.,~][]{vlemmings:06, perez:11}. The latter type of H$_2$O masers are shock-excited and bear a close resemblance in excitation to H$_2$O masers excited toward HMSF, while the former type of H$_2$O masers have an excitation that is significantly affected by the IR continuum, and are not necessarily associated with shocks \citep{gray:16,gray:22}. 

We start by discussing H$_2$O masers that are excited in a shell-like structure, in the circumstellar envelopes of evolved stars. In comparison to SiO masers, most of the H$_2$O maser transitions here are excited farther out from the central star, at about $5-15$ stellar radii \citep{gray:16}. The larger distance from the star, likely diminishes the effect of it on the excitation, as its effect scales with $R^{-2}$, with the star subtending a solid angle of only $\Delta \Omega_* \sim 8\times 10^{-4}$ sr at a distance of $10$ stellar radii. Still, stellar radiation affects the excitation of the maser indirectly, as it heats up the dust to high temperatures. Emission from, in particular optically thin, dust will be highly anisotropic, exhibiting a gradient in the radial direction. The geometry of the maser clumps likely exhibits anisotropy, too, as masers emission are beamed tangentially from the star \citep{richards:12}. From these qualitative arguments, one may expect the anisotropic pumping of these maser sources, but quantitative estimates are needed to confirm this. Full polarization observations by \citet{vlemmings:05b} of $22$ GHz H$_2$O maser toward a sample of late-type stars revealed no detectable linear polarization. 

H$_2$O masers are also observed in association with powerful collimated outflow structures emerging from evolved stars \citep[e.g.,~][]{imai:07}. The masers are associated with shocked regions in the collimated outflow, making our models particularly applicable to these types of masers. Linear and circular polarization have been observed in association with $22$ GHz H$_2$O masers, and have indicated important information on the magnetic field properties of these outflows \citep{vlemmings:06, perez:11}. In these observations, linear polarization degrees of up to a few percents have been observed, consistent with expected polarization degrees of unsaturated, or moderately saturated masers, according to our modeling. These modest linear polarization degrees are not sufficient to explain the observed circular polarization fractions in \citet{vlemmings:06}, thus they provide for a robust magnetic field strength tracer.

Observations of the H$_2$O $22$ GHz maser transition toward high-mass star-forming regions are plentiful and have been a useful tool to map out both the magnetic field morphology, and strength, of star-forming regions \citep{sarma:02, surcis:11a,surcis:11b,surcis:14,goddi:17,surcis:23}. Linearly polarized masers are regularly observed, exhibiting linear polarization fractions between a few tenths of a percent, up to $\sim 25\%$. Linear polarization fractions up to some percents may be explained by anisotropic pumping, and are particularly relevant for unsaturated H$_2$O masers. However, for saturated masers, polarization through maser saturation will likely dominate the polarization signature \citep{surcis:23}. Circular polarization fractions are observed between a few tenths of a percent to some percents. While, occasionally, for the most strongly linearly polarized masers, the degree of linear polarization is sufficient for non-Zeeman effects to be a viable mechanism to explain the circular polarization fraction, for the absolute majority of cases, linear polarization fractions are too low. Thus, also in star-forming regions, $22$ GHz H$_2$O masers are robust tracers of the magnetic field strength.

H$_2$O megamasers are common in the nuclear regions of extragalactic sources, where they are associated with molecular accretion disks and nuclear outflows \citep{lo:05}. The excitation geometry of megamasers is likely anisotropic, as they are excited either in the shocked outflowing gas, or in the Keplerian accretion disk. In addition, water megamasers occur in a busy radiative environment \citep{gallimore:23}. The most studied H$_2$O megamaser is the $22$ GHz maser, but recent work notes the prevelance, circumnuclear association, and high luminosity of the $183$ GHz H$_2$O maser \citep{humphreys:16, hagiwara:21, pesce:23}. While VLBI observations of the $22$ GHz H$_2$O megamaser have provided important information on the intricate kinematics toward galactic nuclear regions \citep[e.g.,~][]{moran:95, kuo:20, gallimore:23}, and provided circumstantial evidence for gas flows that are (partially) magnetically regulated \citep{gallimore:23, kartje:99}, direct detections of magnetic fields, through either linear or circular polarization observations, in H$_2$O megamasers have hitherto been unsuccessful \citep{deguchi:95, herrnstein:98, modjaz:03, vlemmings:07, surcis:20, gallimore:23}. Constraints from non-detections of the circular polarization toward NGC3079, place a limit on the line of sight magnetic field in the strongest maser spot $\lesssim 11$ mG \citep{vlemmings:07}. Non-detections of linear polarization are more difficult to interpret, as even low ionization degrees would cause significant Faraday depolarization of the maser emission \citep{herrnstein:98}.

The observations of maser polarization in H$_2$O masers have been restricted to studies of the $22$ GHz maser, since it is the most common and most luminous of the H$_2$O masers. Also, since it occurs at a rather low frequency, it is the most sensitive probe of the Zeeman effect, that scales inversely with the frequency (see Eq.~\ref{eq:pV_zee}). However, when it comes to the production of linear polarization, we note that we consistently found that the higher frequency $183$ GHz maser is associated with a high degree of anisotropic pumping, and thus expected to be polarized up to about $5\times$ more strongly (in the unsaturated regime) compared to the $22$ GHz maser. In addition, the $183$ GHz transition is significantly less affected by Faraday rotation or depolarization, which is particularly relevant for tracing magnetic fields in megamaser sources. The $183$ GHz maser is commonly associated with late-type stars \citep{yates:95}, protostars \citep{waters:80}, and it also occurs as a megamaser \citep{humphreys:16, pesce:23}. \citet{humphreys:17} find detectable differences in flux density between the two orthogonal polarization receivers of the SEPIA Band 5 receiver on APEX (full calibration of the linearly polarized Stokes parameters could not be achieved as the integration was not obtained over the necessary range of parallactic angles), when observing the $183$ GHz H$_2$O maser toward late-type stars, indicating that the maser is significantly polarized (lower limits of a few percents) while it is likely unsaturated. This is in agreement with our calculations, that indicate that this maser species is likely to be strongly linearly polarized $\sim 10\%$, and may therefore be an excellent tracer of the magnetic field morphology. 

%) toward evolved stars,  where they are  as well as the circumstellar envelopes of evolved stars, such as AGB stars \citep{vlemmings:05}, but also toward collimated outflow structures (water masers toward evolved stars are excited either in a shell like structure around the central late-type star, or in association with a fast and collimated outflow. The latter type of water masers are shock-excited and bear a close resemblance in excitation to water masers excited toward HMSF, while the former type of water masers have an excitation that is significantly affected by the IR continuum, and are not necessarily associated with shocks.)

\subsection{Polarization of class I CH$_3$OH masers}
\label{sec:methanol}
\subsubsection{Linear polarization observations}
\citet{wiesemeyer:04} investigated the linear polarization of millimeter CH$_3$OH masers. They investigated both class I and II masers, and found that while they estimate that saturation levels are modest, polarization degrees are often high. They invoke anisotropic pumping as a mechanism to explain the polarization of unsaturated class II masers, where a directional IR field is the source of the anisotropy. For the class I CH$_3$OH masers, however, they invoke instead collisional polarization \citep{lankhaar:20b} through anisotropic electron collisions, as a means to create anisotropic pumping. Instead of collisional polarization, from our quantitative simulations, it appears more likely that the anisotropic pumping is the result of the shock geometry in which the class I CH$_3$OH masers are excited in. \citet{wiesemeyer:04} investigated the $95$ GHz transition, where they found polarization degrees of $3.8\%$ and $14.5\%$ in two sources. These polarization fractions may be explained by the anisotropic pumping parameters that we obtained. 

In addition, we note that \citet{wiesemeyer:04} find for many of their masers, the polarization fraction adheres to a profile that has low fractional polarization in the line wings and high fractional polarization in the line center. We note that this is a feature that we predict for anisotropically pumped masers, where the polarization fraction, $p_l \sim q_{\mathrm{anis}} \tau_{\nu}$, is proportional to the maser optical depth, which of course in turn adheres to the line profile.

More recently, \citet{kang:16} observed the polarized emission from Class I CH$_3$OH masers toward massive star-forming regions. Focusing on the $44$ GHz and $95$ GHz transitions, that belong to the same family of Class I CH$_3$OH maser transitions, they found that approximately $60\%$ of the sources presented at least some percents of fractional polarization in at least one of the maser species. 

In the sample of \citet{kang:16}, most masers are polarized to degrees $<10\%$. A positive correlation between maser brightness and polarization fraction is present for the strongest masers $>100$ Jy, but it is found that the strongest polarization occurs for the weakest masers, which is likely due to sensitivity effects. While the two masers show similar polarization properties, it is found that the $95$ GHz maser polarizes more strongly compared to the $44$ GHz maser. Just as \citet{wiesemeyer:04} found, the profiles of the linear polarization fraction often follow the line profile, as would be expected for polarization through anisotropic pumping. Thus, while most of the masers in the sample of \citet{kang:16}, are likely unsaturated, they still find high degrees of polarization. Indeed, these polarization fractions may be explained by the mechanism of anisotropic pumping through a shocked geometry. The predicted anisotropic pumping parameters, are an adequate explanation for the observed polarization fractions. In addition, anisotropic pumping can also explain the spectral profile of the linear polarization fraction, that adheres to the profile of the total emission. We predict that a higher polarization fraction of $95$ GHz, compared to $44$ GHz masers, is present when masers are excited in low(er) density gas ($n\sim 10^5$ cm$^{-3}$), thus we suggest this as an explanation for the relatively high degrees of polarization of the $95$ GHz maser observed by \citet{kang:16}.

\subsubsection{Circular polarization observations}
Circular polarization in the $36$ GHz, $44$ GHz and $25$ GHz class I CH$_3$OH masers has been detected using the VLA, and interpreted for its information on the magnetic field \citep{sarma:09,sarma:11,sarma:20}. Circular polarization fractions of $0.06-0.08\%$ were found, indicating magnetic fields in excess of $10$ mG, if interpreted assuming circular polarization through the Zeeman effect and using the Zeeman proportionality constants of \citet{lankhaar:18}.

As discussed in section \ref{sec:lin_circ}, circular polarization may be produced by anisotropically pumped masers if the magnetic field changes direction along the line of sight. When the rotation of the magnetic field projection on the plane of the sky is $45$ degrees, the production of circular polarization is maximal, and on the order of the linear polarization degree squared: $\sim p_l^2$. For weaker rotations, we may roughly scale the conversion $\sim \sin (2\chi) p_l^2$, where $\chi$ is the rotation angle. Then, for a $20$-degree rotation of the magnetic field direction, along the line of sight, the fractional circular polarization that is produced for an anisotropically pumped maser is $p_c \sim 0.3 p_l^2$. Thus, in addition to a modest rotation of the magnetic field, linear polarization fractions on the order of $5\%$ are required to explain the observed circular polarization in class I CH$_3$OH masers. In the typical excitation geometry that we investigated in our modeling, such degrees of linear polarization may easily be produced due to anisotropic pumping. 

Still, the detection of circular polarization is exclusive to only some of the observed masers \citep{sarma:09, sarma:11, sarma:20}, prompting the question of why circular polarization is not more common in these species. Under the hypothesis of circular polarization due to the Zeeman effect, this would be explained by a varying line-of-sight magnetic field strength among the maser spots. Assuming the circular polarization is due to the anisotropic pumping in conjunction with a rotating magnetic field, it could be that other masers are excited in (more) isotropic geometries, or, in particular for the $36$ GHz maser, in denser regions. The only way to unequivocally confirm that the circular polarization is due to the Zeeman effect, is to simultaneously determine the linear and circular polarization of the maser.

\subsubsection{Class II CH$_3$OH masers}
We have not attempted to model class II CH$_3$OH masers. The reason for this is two-fold. First, the excitation of class II CH$_3$OH masers is significantly affected by a dust-phase that is co-spatial with the maser \citep{sobolev:97}. Accounting for the radiative transfer of a co-spatial dust-phase requires additional terms in our radiative transfer models outlined in section \ref{sec:theory}. Second, the excitation of CH$_3$OH masers proceeds through a complicated network of torsionally and vibrationally excited states, that need to be comprehensively modeled in the excitation analysis. While using the formalism we present here does allow for such a comprehensive modeling, it is beyond the scope of this paper. Still, qualitative analysis does suggest that class II CH$_3$OH masers should show a propensity for anisotropic pumping, as (i) fast radiative transitions to torsionally excited states are essential to the maser operation \citep{cragg:05}, (ii) the maser arises in an anisotropic geometry \citep{sobolev:97}, and (iii) class II CH$_3$OH masers are often associated with a nearby HII region \citep{phillips:98}. Class II CH$_3$OH masers will be considered in a follow-up work.
\subsection{Polarization of SiO masers}
Vibrationally excited SiO masers are the prototypical example of an anisotropically pumped maser. The strong linear polarization of the $J=2\to1$ and $J=1\to 0$ transitions, that usually amounts to tens of percents, up to 100 percent, of the total intensity has been observed in many VLBI observations toward late type stars \citep{barvainis:87, mcintosh:89, kemball:97, kemball:09, kemball:11, cotton:11}. ALMA observations have confirmed that high polarization degrees in SiO masers are sustained higher angular momentum transitions \citep{vlemmings:11,vlemmings:17}. 

Estimates suggest that the strongest SiO maser features are moderately saturated, $R \lesssim 10 \bar{\gamma}$ \citep{tobin:19, vlemmings:17}. Expected polarization levels through saturation polarization under these conditions are on the order of some percents. Thus, for the masers that are polarized in excess of $\sim 10\%$, anisotropic pumping needs to be invoked to explain these high linear polarization yields. Indeed, our quantitative simulations confirm that SiO masers are prone to produce polarization through their anisotropic pumping, and may explain polarization degrees up to $100\%$.

The degree of anisotropic pumping is a strong function of the influence of the central star on the pumping of the masers. However, the relative contribution of the infrared emission of the central star is still a matter of debate \citep{lockett:92, gray:09, gray:12}. Still, we do expect anisotropic pumping of SiO masers, also when they are collisionally pumped, since the decay of the vibrationally excited rotational, through ro-vibrational transitions, is the main mechanism through which population inversion is achieved. It is well-established that SiO masers are tangentially beamed \citep{kemball:97, gray:12}, which thus also indicates anisotropy in their radiative relaxation. This will subsequently manifest as anisotropic pumping, albeit at more modest degrees compared to situations where the infrared radiation field from the central star impacts the excitation significantly.

SiO masers are also regularly observed to exhibit circular polarization; a feature which has been used to derive information about the magnetic field strength in the extended atmosphere of evolved stars. The most comprehensive sample of SiO circular polarization observations comes from \citet{herpin:06}, who used IRAM to observe SiO masers toward numerous sources in full polarization mode. SiO masers were observed to regularly exhibit circular polarization, mostly on the order of some percents. Assuming this circular polarization is due to the Zeeman effect, this yields magnetic field strengths on the order of some Gauss. The observations of \citet{herpin:06} also included linear polarization, which generally was observed to be stronger than the circular polarization. Interestingly, a linear regression fit related the circular polarization to the linear polarization as: $p_c[\%] \approx 1.5\% + 0.25 p_l[\%]$. It should be noted, however, that linear polarization directions may vary within SiO maser clumps \citep{kemball:11}, which would lead to an underestimation of the actual linear polarization fraction when such masers observed with a single dish telescope.

In particular for SiO masers, that are anisotropically pumped, a changing magnetic field along the line of sight will lead to the production of circular polarization. Indeed, changing polarization angles, that indicate changing magnetic field directions, have been observed within SiO maser clumps \citep{kemball:11, tobin:19}. We estimated that for small changes in the magnetic field angle, circular polarization will be produced on the order of $p_c \sim p_l^2$. For a significant fraction of the masers in the sample of \citet{herpin:06}, this mechanism is sufficiently effective to produce the observed circular polarization fractions. Still, for the largest part of the sample $p_c > p_l^2$ and a Zeeman origin of the circular polarization appears the most viable explanation, but we do note that for those masers linear polarization fractions are possibly underestimated due to resolution effects. In particular for SiO masers, to extract information about the magnetic field strength from circular polarization observations, it is of importance to rule out non-Zeeman circular polarization mechanisms for each maser clump under consideration.

\subsection{Linear polarization through the Zeeman effect}
\label{sec:zee_pol}
Finally, it is interesting to compare the production of linear polarization through anisotropic pumping to the production of linear polarization through the Zeeman effect. Due to the spectral decoupling of the $\Delta m=\pm 1$ and $\Delta m =0$ transitions, that are associated with different opacities for the polarization modes of the radiation field, polarization is produced through the Zeeman effect. Most commonly, the signature of the Zeeman effect is sought in the circular polarization, as the circular polarization is linearly proportional to the Zeeman effect, which is in turn proportional to the magnetic field strength. However, also linear polarization and line broadening are produced through the Zeeman effect \citep{lankhaar:23}, which are features that are quadratically proportional to the magnetic field strength. Using the approximations outlined in \citet{lankhaar:23} and Chapter 9 of \citet{landi:06}, and focussing on the propagation of (polarized) radiation toward the line center, where circular polarization approaches zero, we note the polarized propagation coefficients due to the Zeeman effect
\begin{subequations}
\label{eq:prop_zee}
\begin{align}
\kappa_{\parallel,\perp}^{\mathrm{Z}}(\nu,\Omega) &= \kappa_I^{\mathrm{Z}}(\nu,\Omega) \pm \kappa_Q^{\mathrm{Z}}(\nu,\Omega), \\ 
\kappa_I^{\mathrm{Z}}(\nu,\Omega) &= k_0\left[\phi_{\nu} +\nu_Z^2 \left(\frac{\bar{Q}}{4} - \frac{\Delta Q}{4} \cos^2 \vartheta \right) \frac{d^2 \phi_{\nu}}{d \nu^2}  \right] \nonumber \\ &\simeq k_0 \phi_{\nu} \\
\kappa_Q^{\mathrm{Z}}(\nu,\Omega) &= k_0 \left[\nu_Z^2 \frac{\Delta Q}{4} \frac{d^2 \phi_{\nu}}{d \nu^2} \sin^2 \vartheta  \right],
\end{align}
\end{subequations}
where $\nu_Z$ is the average Zeeman shift of the $\Delta m=\pm 1$ transition groups in frequency units and $\bar{Q}$ and $\Delta Q$ are dimensionless coefficients that derive from the intragroup ($\Delta m=0,\pm 1$) spread in Zeeman shifts \citep{lankhaar:23, landi:06}. We note that $\bar{Q}$ is positive, while $\Delta Q$ may either be positive or negative. The Zeeman shift $\nu_Z = z B / 2$ is related to the magnetic field by the Zeeman factor, $z$, which on the order of $\mathrm{Hz}/\mathrm{mG}$ for non-paramagnetic molecules. For a $J=1-0$ transition, or for linear molecules, the dimensionless parameters $\bar{Q}=-\Delta Q = 1$, as for these transitions, no intra-group Zeeman shift spread is present. But for transitions with higher angular momentum, the factors $\bar{Q}$ and $\Delta Q$ can assume values in excess of $1000$ \citep[see discussion in ][]{lankhaar:23}. %We list the relevant Zeeman coefficients for well-known maser transitions in Table \ref{tab:zeeman_params}. 
The simplification associated with the $\kappa_I^{\mathrm{Z}}$ is warranted when the Zeeman shift is significantly weaker than the line width. 

We are now in a position to study the transfer of polarized radiation toward the line center of a Zeeman splitted maser transition. At resonant frequency, assuming a Doppler profile with full width at half maximum (FWHM) $\Delta \nu_{\mathrm{FWHM}}$, we may set $\left.d^2 \phi_{\nu}/d\nu^2 \right|_{\nu_0}=-8 \log 2 / \Delta \nu_{\mathrm{FWHM}}$. Then, using the radiative transfer equation of Eq.~(\ref{eq:rad_trans}) with propagation coefficients Eq.~(\ref{eq:prop_zee}) at resonant frequency, assuming an unsaturated maser, we may compute the linear polarization fraction due to the Zeeman effect,
\begin{subequations}
\label{eq:pQ_zee}
\begin{align}
p_Q^{\mathrm{Z}} &= \tanh{ \left[- \tau_{\nu} \frac{2 \log 2 \ \nu_Z^2 \Delta Q}{\Delta \nu_{\mathrm{FWHM}}^2} \sin^2 \vartheta \right]}  \nonumber \\ &= \tanh{ \left[- \tau_{\nu} q_{\mathrm{Z}} \sin^2 \vartheta \right]}.
\end{align}
We note that production of polarization for unsaturated masers, both through the Zeeman effect and through anisotropic pumping, have a similar dependence on the optical depth and projection angle onto the magnetic field, $\vartheta$. In fact, the linear dependence (for low polarization degrees) of the polarization fraction on the optical depth had already been found in the proper polarized radiative transfer modeling of \citet{tobin:23}, and is neatly captured in our analytical model. We compute the dimensionless parameter, $q_{\mathrm{Z}}$, that is akin to $q_{\mathrm{anis}}$, for a non-paramagnetic molecule under typical interstellar conditions
\begin{align}
q_{\mathrm{Z}} &= 1.2 \times 10^{-5} \left(\frac{z}{1 \ \mathrm{Hz/mG}} \right)^2 \left(\frac{B}{100 \ \mathrm{mG}} \right)^2 \nonumber \\ &\times \left( \frac{\nu_0}{10 \ \mathrm{GHz}}\right)^{-2} \left(\frac{\Delta v_{\mathrm{FWHM}}}{1\ \mathrm{km/s}} \right)^{-2} \Delta Q.
\end{align}
\end{subequations}
As a final note, it may be shown that the Zeeman circular polarization of an unsaturated maser, is $p_V^{\mathrm{Z}}  = 2 \sqrt{2 \log{2}/e} \frac{\nu_Z}{\Delta \nu_{\mathrm{FWHM}}} \cos \vartheta$. This relates the linear polarization to the circular polarization (for polarization $<10\%$) of unsaturated masers, which are polarized by the Zeeman effect, by
\begin{align}
p_Q^{\mathrm{Z}} &\simeq - \tau_{\nu} \frac{2 \log 2 \ \nu_Z^2 \Delta Q}{\Delta \nu_{\mathrm{FWHM}}^2} \sin^2 \vartheta \nonumber \\ &= -\tau_{\nu} \frac{e}{4} \left(p_V^{\mathrm{Z}}\right)^2 \Delta Q \tan^2 \vartheta. 
\end{align}
The detection of linear polarization in an unsaturated maser, which is polarized though the Zeeman effect, is thus expected to be strongly associated with the presence of circular polarization. Both the circular ($\propto B\cos \vartheta$) and linear polarization ($\propto \tau_{\nu_0} B^2 \sin^2 \vartheta$, with polarization angle $\chi = \frac{1}{2} \arctan{U/Q}$ indicating the magnetic field direction in the plane-of-the-sky) properties may then be used to acquire the 3D magnetic field direction inside the maser region. In practice, as non-paramagnetic masers are rarely found to be circularly polarized in excess of some percents, their linear polarization due to the Zeeman effect is expected to be low. This technique may be applied, however, to paramagnetic maser species such as OH.

\section{Conclusions}
We presented excitation modeling of masers species, where we explicitly resolved the magnetic sublevels when solving for the quantum state populations. In an anisotropic geometry, transition rates between magnetic sublevels are a function of the geometry, and its relation to the magnetic field direction \citep{goldreich:81}. Only in the case of an isotropic geometry will the transition rates between different magnetic sublevels of the same transition between two rotational states be equal. To avoid a drastic increase in the dimensionality of the excitation analysis, we used an irreducible tensor formalism, that groups magnetic sublevels on the basis of their transformation properties. This procedure is highly advantageous as it allows for a dimension reduction, to only twice that of a regular excitation analysis, at minimal, $<1\%$, expense of accuracy \citep{lankhaar:20a}. When a maser excitation analysis is performed in this way, the anisotropic pumping parameters can be extracted.

We presented excitation modeling for H$_2$O masers, class I CH$_3$OH masers, and SiO masers. H$_2$O masers and class I CH$_3$OH were modeled in typical shocked geometries, while SiO masers were modeled including an anisotropic radiation infrared radiation field, that is representative of the environment of SiO masers excited in the extended atmosphere of evolved stars. From our excitation modeling, we extracted the anisotropic pumping parameters. These represent the first quantitative estimates of the degree of anisotropic pumping. The (anisotropic) pumping parameters were subsequently used in full polarized maser radiative transfer simulations to estimate the effect of anisotropic pumping on the polarization of (un)saturated masers.

We investigated the $22$ GHz, $183$ GHz and $321$ GHz H$_2$O maser transitions. We find that the $22$ GHz and $321$ GHz transitions are marginally anisotropically pumped. Anisotropic pumping of these masers can explain polarization degrees of up to some percents in unsaturated masers, and in saturated masers, the polarization signature is dominated by regular saturation polarization. The $183$ GHz maser exhibits moderately high levels of anisotropic pumping, which can produce polarization fractions up to $10\%$ in unsaturated masers, and impacts either enhancing or diminishing, the production of polarization due to maser saturation, significantly. 

The polarization properties of class I CH$_3$OH masers have been investigated for the $25$ GHz, $36$ GHz, $44$ GHz and $95$ GHz transitions. We find that the $25$ GHz and $36$ GHz transitions are modestly anisotropically pumped. Anisotropic pumping for these masers can explain polarization degrees on the order of $5\%$ in unsaturated masers. The $44$ GHz and $95$ GHz transitions exhibit significant anisotropic pumping, that can lead to polarization degrees of $15\%$ in unsaturated masers, and will impact the production of polarization in saturated masers significantly. Our simulations can explain the observed polarization fractions in linear polarization observations of class I CH$_3$OH masers by \citet{wiesemeyer:04} and \citet{kang:16}.

SiO masers are found to be highly anisotropically pumped in all their vibrationally excited maser transitions. The anisotropy of the pumping is mainly due to the directional radiation from the central star, but is also expected to be present in fully collisionally pumped masers due to anisotropic radiative decay through vibrational relaxation transitions. Our quantitative estimates of the anisotropic pumping parameters can justify polarization degrees up to $100\%$.

Regardless of the origin of the maser polarization, coming either from maser saturation or anisotropic pumping, the polarization direction is either parallel or perpendicular to the magnetic field direction projected on the plane of sky. This is regardless of the polarization mechanism, as long as the magnetic precession rate exceeds the rate of stimulated emission: $R \ll g\Omega$.

We outlined a mechanism through which an anisotropically pumped maser can produce non-Zeeman circular polarization when the magnetic field along the line-of-sight changes direction. The non-Zeeman circular polarization fraction, $p_c \sim p_l^2$, is quadratically related to the linear polarization fraction. In contrast to other non-Zeeman circular polarization mechanisms \citep{nedoluha:94, wiebe:98}, the mechanism we propose operates at arbitrary degrees of saturation.

\begin{acknowledgements}
BL acknowledges support for this work from the Swedish Research Council (VR) under grant number 2021-00339. The simulations were enabled by resources provided by Chalmers e-Commons at Chalmers. We thank the referee for comments that improved the quality of the paper.
\end{acknowledgements}

%Recently, with the advent of ALMA, (sub-)millimeter masers have become available for interferometric studies, too.

%n_H2 = 1e8,1e9,1e10
%\epsilon=0.01,0.1,10,100
%\Xi bepaald de range voor N_H2/\Delta v -- kijken naar NM91 voor een goeie range
%log_10(\xi)=[-2,2]
%\Delta_{FWHM}=1 km/s / 2 km/s (megamasers)
%nH2 = 1e9          log_10(N_spec) = 1e16 cm-2 / km/s - 1e20 cm-2 / km/s
%1e8                log_10(N_spec) = 1e17 cm-2 / km/s - 1e21 cm-2 / km/s
%1e10               log_10(N_spec) = 1e15 cm-2 / km/s - 1e19 cm-2 / km/s
%T = 500 Kelvin

%transitions of interest
%22 GHz
%183 GHz
%321 GHz

%\subsubsection{Star forming regions}
%n_H2 = 1e8,1e9,1e10
%\epsilon=0.01,0.1,10,100
%\Xi bepaald de range voor N_H2/\Delta v -- kijken naar NM91 voor een goeie range
%log_10(\xi)=[-2,2]
%\Delta_{FWHM}=1 km/s / 2 km/s (megamasers)
%nH2 = 1e9          log_10(N_spec) = 1e16 cm-2 / km/s - 1e20 cm-2 / km/s
%1e8                log_10(N_spec) = 1e17 cm-2 / km/s - 1e21 cm-2 / km/s
%1e10               log_10(N_spec) = 1e15 cm-2 / km/s - 1e19 cm-2 / km/s
%T = 500 Kelvin

%\subsubsection{Megamasers}
%Add cold dust and increase column densities

\bibliography{lib.bib}

%log10(N_mol) = 14-17 cm2 km-1 s 
%nH2 = 1e7
%T=100 K
%\epsilon=0.01,0.1,10,100

%transitions of interest
%E-type
%25 GHz 3_2 - 2_1
%36 GHz 4_-1 - 3_0

%A-type
%44 GHz 7_0 - 6_1 
%
\end{document}